\documentclass[12pt]{iopart}
\usepackage[utf8]{inputenc}
\usepackage{graphicx}
\usepackage{color}
\begin{document}
\title{The role of low-energy electrons in the charging process of LISA test masses}
\author{Simone Taioli$^{1,2,3}$, Maurizio Dapor$^{1,2,*}$, Francesco Dimiccoli $^{2,4}$, Michele Fabi $^{5,6}$, Valerio Ferroni$^{2,4,*}$, Catia Grimani $^{5,6}$, Mattia Villani$^{5,6}$, William Joseph Weber$^{2,4}$}
\address{$^1$European Centre for Theoretical Studies in Nuclear Physics and Related Areas (ECT*), Bruno Kessler Foundation, Trento, Italy\\
$^2$Trento Institute for Fundamental Physics and Applications (TIFPA-INFN), Trento, Italy\\
$^3$Faculty of Applied Physics and Mathematics, Gdańsk University of Technology, Gdańsk, Poland\\
$^4$Department of Physics, University of Trento, Italy\\
$^5$Università degli Studi di Urbino Carlo Bo, Via S. Chiara 27, 61029, Urbino (PU), Italy\\
$^6$ Istituto Nazionale di Fisica Nucleare, Sez. di Firenze, Via Bruno Rossi, 1, 50019, Sesto Fiorentino, Firenze, Italy\\
$^*$ corresponding author}



\begin{abstract}
The estimate of the total electron yield is fundamental for our understanding of the test-mass charging associated with cosmic rays in the LISA Pathfinder mission and in the forthcoming gravitational wave observatory LISA. To unveil the role of low energy electrons in this process owing to galactic and solar energetic particle events, in this work we study the interaction of keV and sub-keV electrons with a gold slab using a mixed Monte Carlo and ab-initio framework. We determine the energy spectrum of the electrons emerging from such a gold slab hit by a primary electron beam by considering the relevant energy loss mechanisms as well as the elastic scattering events. We also show that our results are consistent with experimental data and Monte Carlo simulations carried out with the GEANT4-DNA toolkit.

\end{abstract}

\submitto{\CQG}

\section{Background and motivations}

The realization of drag-free spacecraft has opened new frontiers to space exploration as well as to missions aiming at testing the laws of fundamental physics \cite{Lange,TRIAD,Toubul,GPB_main,LPF_main,PhysRevLett.129.121102}. A drag-free spacecraft hosts a proof-mass in free fall shielded from external forces, such as drag and solar pressure, as a geodesic reference system for its trajectory. The spacecraft that follows its free falling proof-mass traces an orbit which contains valuable information of the local gravitational field useful for geodesic applications.
On the other hand, the drag-free satellite proof-mass is a practical realization of a local inertial reference frame, and thus can be used as a test body to explore the fundamental laws of physics. Proof-masses are used in the LISA mission as both free-falling references of geodesic motion and end-mirrors for an interferometric measurement of the gravitational wave strain. It is worthwhile to point out that the proof-masses are in pseudo free fall because of the influence of the local thermal, gravitational and electromagnetic environment determined by their own satellite. To limit these disturbing forces they need to be electrically neutral with respect to their housing. Some missions ground the proof-masses using a thin gold wire, at the cost of additional force noise (damping $\approx f^{-1/2}$) that limits the free fall performance at the lowest frequencies \cite{Toubul,PhysRevLett.129.121102}, and will not be compatible with LISA observatory  performance goal. GP-B was the first to use UV light and the photoelectric effect to control the charge on its gyroscopes avoiding any mechanical contact \cite{GPB_main}. Other missions afterwards, used or planned to use photoelectric devices for the proof-mass charge management \cite{LPF_main,Saraf_2016,LISA,DECIGO,Taiji,Tianqin}. The performance of these systems relies on the accurate knowledge of the space environment and its interaction with the spacecraft.  In the space environment, highly energetic particles of galactic and solar origin easily penetrate the light structures of the spacecraft leading to the charging of the proof mass. The energy spectra of primary and secondary particles reaching the proof masses span over several orders of magnitude. Thus, in this work we focus on the role played by keV and sub-keV electrons in the charging process of the LISA proof-masses.

\section{Introduction}
\label{s1}
The Laser Interferometer Space Antenna (LISA) \cite{LISA,Bender} will be the first gravitational wave detector in space to unveil the secrets of the mHz gravitational universe. The mission, 
designed and led by ESA with NASA participation, is in its phase B1 and it is scheduled for launching in $2034$. LISA will consist of three spacecraft orbiting the Sun in a triangular constellation. Each spacecraft hosts a pair of free falling proof-masses (TMs) following geodesic motion to reveal the tidal acceleration of gravitational waves passing through the satellite constellation. The ESA LISA Pathfinder (LPF) \cite{LPF_main} mission tested most of the technology necessary for LISA, in particular that it is possible to place a TM in geodesic motion in space with the required sub-femto-g residual acceleration noise in the frequency band $0.1$—$40\,\mathrm{mHz}$ \cite{LPF_PRL1,LPF_PRL2}. 

The TMs are 2 kg Au/Pt cubes surrounded by an electrode housing (EH) separated by a vacuum gap ranging from 2.9 to 4 mm \cite{WeberSPIE,Dolesi2003}. The EH provides the conducting shield from external electromagnetic disturbances, the TM sensing and actuation, and also the local ground potential reference for the TMs. In science operations, voltages of the order of $1$ V are applied to the electrodes. 
TMs as EHs are gold-coated with no mechanical or electrical contact to the spacecraft, and as a consequence they accumulate charge \cite{ara05} due to the cosmic rays of galactic \cite{grim15} and solar \cite{grimani2014} origin interacting within the spacecraft. Any stray electrostatic field, originated from patch effects \cite{PE1,PE2} or from applied actuation voltages, couples to the TM charge and produces forces and force gradients \cite{PE3}. Moreover, cosmic-ray charging fluctuates\footnote{ given the Poissonian nature of the charging process and possible additional low-frequency noise due to the modulations of the interplanetary galactic cosmic-ray (GCR) flux.} as do stray electrostatic fields.   
Noise in both TM charging and fields leads to important contributions to the TM force noise measured in LPF, which is relevant for the LISA acceleration noise budget  \cite{LPF_Peter,Bill_PRL,Sumner2022,shaul2005}.
Additionally, GCR short-term variations can be responsible for coherent charge-induced forces that can be confused with transient gravitational wave signals in the LISA science data. 
Even in the absence of stray electrostatic potentials, the TM charge creates a force gradient, depending quadratically on the amount of deposited charge, which couples the TMs to the noisy spacecraft motion. This creates additional force noise and affects the TM control. 
As a result, TM charging  must be limited for LISA. The TM discharging system will consist of Ultra-Violet (UV) illumination to generate photo-electron currents from the TM and the EH surfaces that allows to bring the TM to the desired potential level. This technology was inherited from GP-B mission \cite{GPB_main,GPB_Buchman} and tested successfully in LISA Pathfinder \cite{LPF_Dan}. In LPF, the TMs were discharged with UV Hg lamps, while in LISA a UV Light Emitting Diode (LED) based system is devised \cite{UF1}.

The discharging process was carried out either periodically, with the interruption of science operations, or continuously. The latter approach induced a small but continuous noise to the TM motion.  

It is of primary importance to study the TM charging process before the mission launch in order to design the charge control system and to consolidate the budget of the charge-induced force noise for LISA. The LISA TM charging is associated with energetic particles able to penetrate about $16\,\mathrm{g\ cm}^{-2}$ of shielding material surrounding the TMs\footnote{We considered about $20\%$ of additional material grammage with respect to LISA Pathfinder on the basis of the current information of the LISA spacecraft design \cite{bridging} }. This sets a minimum energy threshold  of $100\,\mathrm{MeV}$ to the energy of hadrons of galactic and solar origin penetrating or interacting in the spacecraft. This limit decreases to $20\,\mathrm{MeV}$ for electrons and positrons.
Monte Carlo (MC) simulations of the LPF TM charging carried out with FLUKA \cite{Fluka} and GEANT4 \cite{Geant4} before the mission was launched at the end of 2015 returned very similar results for TM charging rate and noise at solar minimum conditions \cite{ara05,grim15,wass2005}, despite different nominal energy limits for electron propagation in GEANT4 ($250\,\mathrm{eV}$) and FLUKA ($1\,\mathrm{ keV}$). This can be explained in terms of the average ionization potential in gold that sets the actual limit of hadron ionization to 790 eV in GEANT4 \cite{ara05}.
The input cosmic ray fluxes for the FLUKA simulations carried out just before the mission launch \cite{grim15}  were predicted on the basis of the  solar modulation expected for the first part of the LPF mission operations. The comparison of these last simulation outcomes with the LPF measurements, dated April 2016 \cite{LPF_Peter}, showed that the net charging was roughly $+25\,\mathrm{e/s}$, in the middle of the range predicted before flight, while the observed charge noise was 3-4 times larger than the estimates. 
In addition, we observed that the charging rates were within the expected range but measurably different on the two TMs\footnote{LISA Pathfinder hosted a couple of TMs and EHs for differential acceleration measurements: one TM was used as geodesic reference for the satellite, the other one as reference inertial body to measure the local forces of deviation from the pure free-fall motion.}.
The difference in the charge rates may have originated from the
different volt-scale AC electrostatic fields used for force
actuation \cite{LPF_Peter}. 
The possible causes of these observations -- namely an increase in charge fluctuations without any significant effect on the net charging, and a dependence on rather small electrostatic potentials --  
were carefully evaluated.  On one hand, a certain fraction of electrons emitted by the surfaces turns out having a velocity parallel to the electric field small enough to be sensible to the potential barrier between the TM and the EH surfaces. On the other hand, in particular, the observations were suggesting that a large amount of charges of the same sign were both entering and escaping the TMs. This scenario is consistent with low-energy electrons (LEE) 
emitted roughly uniformly by the outer layers of the TM and EH Au surfaces traversed by  high-energy particles \cite{ara05}. It is worthwhile to recall that keV electron propagation in materials such as gold is of order of microns, setting this limit to the layer of gold relevant to electron emission that can influence TM charging.
Moreover, LEE are mostly free to propagate in the gap between the TM and the EH where potentials of the order of $1\,\mathrm{V}$  are found. 
The discrepancy between the predictions of MC simulations and experimentally observed LPF TM charging 
noise measurements recorded during the mission operations in April 2016 have been recently bridged by using a dedicated program including low-energy electromagnetic processes (10-1000 eV), called LEI, in addition to FLUKA \cite{mattia,grimcqg21,bridging}. 

These findings motivated us to explore low-energy physics electromagnetic processes by using a more comprehensive approach to the electron transport in gold via a tailored transport MC method and code (SEED, Secondary Electron Energy Distribution) \cite{taioli2010electron,taioli2009mixed,taioli2009surprises,pssb.200982339,AZZOLINI2017299,azzolini2018anisotropic,azzolini2018secondary,taioli2020relative,pedrielli2021electronic,de2022energy,Dapor_book_blu}.

The present work is also motivated to extend to GEANT4 a simulation work for LISA and, in general, for 
applications of LISA gravitational reference sensor hardware to geodesy missions \cite{WeberGeo,UFLGeo} or for 
other proposed gravitational wave observatories \cite{DECIGO,Taiji,Tianqin}. We note that this topic has been considered of great relevance for space mission design by ESA and is the object of a current agency-funded study \cite{ESAITT}.

SEED is a stochastic method that can be used to model the motion of electrons of several hundreds eV energies  impinging at a given angle on a target surface and traveling thereafter inside the specimen. In a typical experimental set-up the emitted primary and secondary electrons (even though this distinction is somehow forced, the electrons being indistinguishable particles) are collected and counted by a spectrometer covering a large or the entire solid angle.
Within the SEED framework, the classical trajectory of each electron of the primary beam can be tracked, from the entrance into the target to the escape out of the solid (or its capture if the charge succeeds to completely deposits its energy), provided its motion satisfies given energy and angular conditions.
In this respect, SEED relies on the accurate determination of the elastic and inelastic scattering cross sections. 
The electrons traveling within the material make stochastic hops according to the probability distribution provided by such interactions. 

Elastic scattering events occur typically when the electrons interact with the constituent atomic nuclei of the sample emerging at some deflection angle with unvaried kinetic energy owing to their mass difference.
Inelastic scattering events instead are mainly due to electron-electron interaction with an electron energy loss implied.
In this regard, several energy-loss mechanisms may occur, such as the excitation and the ionization of the target atoms, which lead to generating secondary electrons (including the Auger electrons). The information on how electrons lose energy during their motion within the solid target is embedded in the so-called energy loss function (ELF). This function, which can be derived from experimental measurements of electron energy loss spectra (EELS)  \cite{Tougaard2004} in transmission (TEELS) or reflection (REELS) geometry, is a unique fingerprint of the material and, properly weighted and integrated, gives access to the inelastic mean free path (IMFP) \cite{taioli2010electron,taioli2009mixed,taioli2009surprises,AZZOLINI2017299,azzolini2018anisotropic,azzolini2018secondary,taioli2020relative,pedrielli2021electronic,de2022energy,Ritchie_PhysRev_1957}. To obtain a complete and accurate description of the charge transport within a solid target, so as to assess the electron energy spectra and the total yield (TEY), defined as the ratio between the reflected and the incident particles on target, the contribution of both primary and secondary electrons must be included in the simulation \cite{Dapor_book_blu}.

The manuscript is organized as follows: in Section \ref{s2} we present the calculation of the electron scattering cross sections for gold, the material used to coat the TMs and their electrode housings. We finally provide a detailed numerical estimate of the LEE emitted from the gold surfaces by using the SEED approach.
In Section \ref{s3} we first show the consistency between the electron transport estimated in Sect. \ref{s2} with dedicated MC simulations performed by GEANT4-DNA (ver 11.0). Finally, we discuss the impact of the LEE onto the LISA TM charging. 

\section{The SEED algorithm}
\label{s2}
 
In our MC algorithm implemented in the in-house software SEED, the occurrence of either elastic or inelastic scattering is evaluated by comparing random numbers with the relevant probability distributions.
These probability distributions are obtained from the knowledge of the elastic and inelastic scattering cross sections, computed via the Mott theory \cite{mott1929scattering} and the Ritchie dielectric theory \cite{Ritchie_PhysRev_1957}, respectively. 
In the next sections we will review thoroughly the theoretical and computational concepts of our MC framework.

\subsection{Kinetic energy distribution of the primary beam}

The initial kinetic energy of the monoenergetic electron beams impinging on the solid target can be determined by analyzing the experimental elastic peak, which collects the back-scattered electrons that underwent zero-loss or quasi-elastic (phonon emission and absorption) scattering.
In particular, the elastic peak shows an energy distribution around the maximum at $E_0$, owing to the finite resolution of the spectrometer and the energy width of the initial beam characterised by a full width ${\Delta E}$, which can be modeled by a Gaussian or a Lorentzian function. The area below the elastic peak, by which we typically rescale the spectral lineshape, can be used to normalize the cumulative distribution probability 
\begin{equation}\label{cumini}
  P(E')= \frac{1}{Area}\int\limits_{E_{-}}^{E'} f(E) d(E),
\end{equation}
where $f(E)$ is the (Gaussian, Lorentzian) function that describes the elastic peak distribution and 
\begin{equation}
   Area = \int\limits_{E_{-}}^{E_{+}} f(E) d(E).
\end{equation}
with $E_\pm=E_0\pm \Delta E$.
For each injected electron of the beam, the value of the initial kinetic energy is determined by generating a uniformly distributed random number $\mu_1$ in the range $[0,1]$ that equates to the cumulative probability of Eq. \ref{cumini}. The electron kinetic energy is increased by the material work function ($\Phi$) after penetrating the surface.

\subsection{Elastic scattering}

\begin{figure}[htp!]
    \centering
    \includegraphics[width =0.8\linewidth]{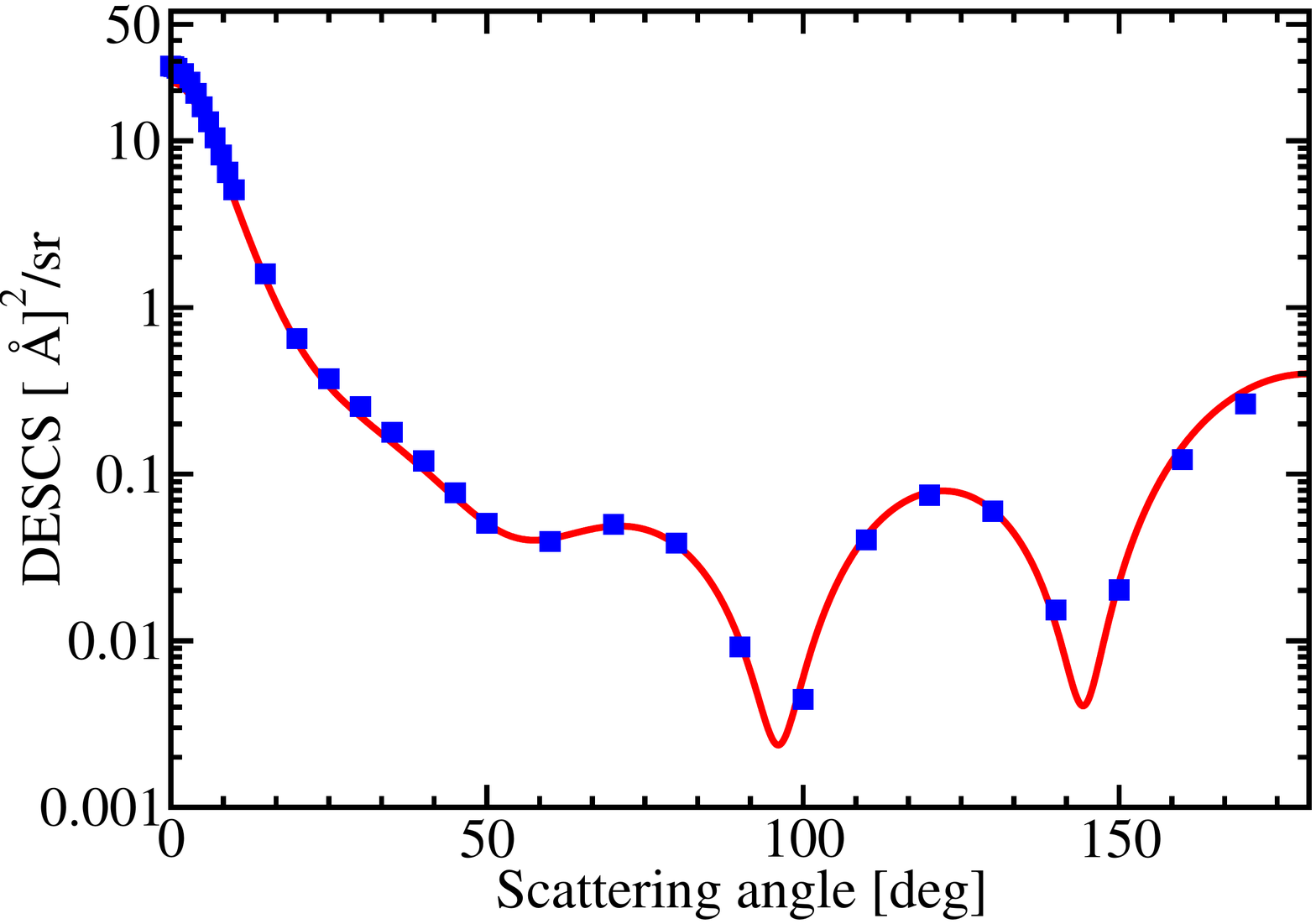}
    \includegraphics[width=0.8\linewidth]{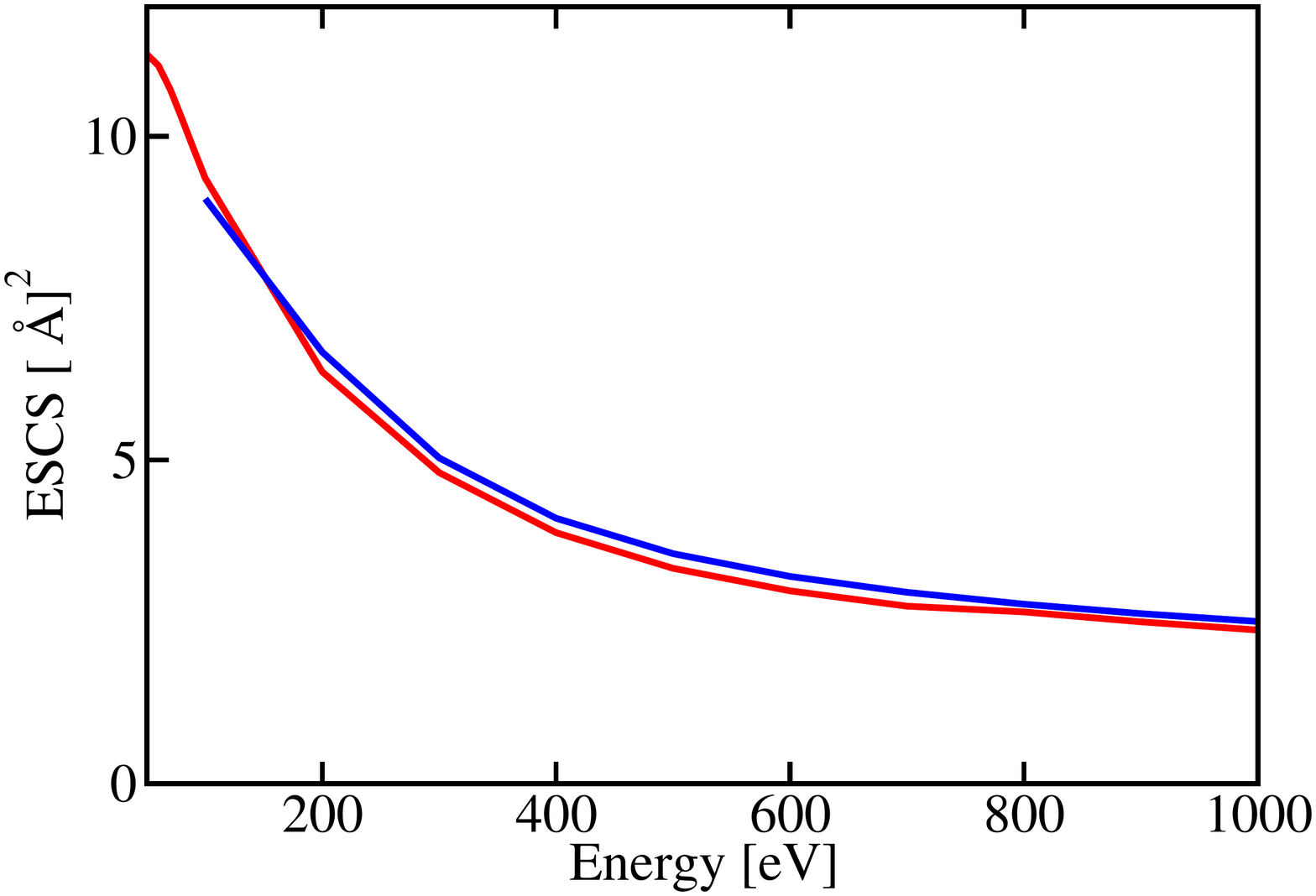}
    \caption{Top: differential elastic scattering cross section of Au at beam kinetic energy of 1000 eV calculated using Eq. \ref{eq:descs} (red line) in comparison to calculations by Riley et al. \cite{RILEY1975443} (blue full squares). Bottom: total elastic scattering cross section of Au as a function of the electron kinetic energy obtained by integrating Eq. \ref{eq:descs} in the solid angle (red line) in comparison to Mayol-Salvat calculations (blue line) \cite{MAYOL199755}.}
    \label{fig:escs}
\end{figure}

The elastic scattering between electrons and the atomic nuclei of the solid target can be described by either solving the Dirac equation in a central field using some approximation to treat the exchange-correlation interaction, such as in the Dirac-Hartree-Fock (DHF) approach \cite{taioli2020relative,de2022energy,taioli2021relativistic}, or the Mott theory \cite{mott1929scattering}, by which the differential elastic scattering cross section (DESCS) can be obtained as:
\begin{equation}
    \frac{d\sigma_\mathrm{el}}{d\Omega} = |f|^2 + |g|^2,
    \label{eq:descs}
\end{equation}
where $f$ and $g$ are the direct and spin-flip scattering amplitudes, respectively \cite{Dapor_book_blu,jablonski2004comparison}. In the latter formulation, which treats the constituents atoms essentially as independent scattering centers (at least concerning the scattering potential), the DHF screening interaction can be modelled by a linear combination of Yukawa potentials and adds multiple scattering effects by introducing a phase factor proportional to the interatomic bond equilibrium distances \cite{taioli2020relative,de2022energy}. The Yukawa potential parameters for Au are taken from Ref. \cite{PhysRevA.36.467},

Finally, the total elastic scattering cross section (ESCS) $\sigma_\mathrm{el}(E)$ can be obtained by integrating Eq. (\ref{eq:descs}) in the solid angle at each kinetic energy $E$. We show the results (red lines) in the top (DESCS) and bottom (ESCS) panels of Fig. \ref{fig:escs} in comparison to the calculations by Riley et al. \cite{RILEY1975443} (blue full squares) and Mayol and Salvat \cite{MAYOL199755} (blue line), respectively.

The elastic scattering mean free path (EMFP) is thereafter calculated, at a given kinetic energy, as: 
\begin{equation}
    \lambda_{\mathrm{el}}(E) = \frac{1}{\cal{N} \sigma_\mathrm{el}(E)},
\end{equation}
where $\cal{N}$ is the atomic density of the target material.

In SEED simulations, the elastic scattering events lead only to a change in the direction of the electron path. 
Our MC algorithm proceeds thus to evaluate the angular deflection of the electron trajectory with respect to the incident direction. The scattering angle, in particular, is obtained by assessing the cumulative elastic scattering probabilities for several different values of the electron kinetic energy $E$: 

\begin{equation}\label{cumel}
    P_\mathrm{el}(\overline{\theta}, E) = \frac{2\pi}{\sigma_\mathrm{el}(E)}\int\limits_0^{\overline{\theta}} \frac{d\sigma_\mathrm{el}(E)}{d\Omega} \sin\theta d\theta,
\end{equation}

\noindent and by equating the integral in (\ref{cumel}) to a random number $\mu_2$ uniformly distributed in [0,1]. 

\subsection{Inelastic scattering}

During their path within the solid, electrons can transfer all or a fraction of their energy and momentum to the atomic constituents of the sample, which may result in (resonant) excitation or ionization processes. 
 These mechanisms can be dealt with by using two different strategies \cite{Dapor_book_blu}: on the one hand, the continuous slowing-down approximation, where the charge loses its kinetic energy continuously over its path; on the other hand, the energy straggling approach, which takes properly into account the energy loss mechanisms, including the possibility that the electrons may lose the whole of their energy in only one inelastic collision. Since our goal is to accurately assess the energy loss spectrum and the TEY, we will use the latter approach that is able to  describe the statistical fluctuations of the different energy loss mechanisms, such as the secondary cascade. We notice that integrated quantities, such as the TEY, are less affected by the choice of the algorithm for treating the energy loss. In this respect, a key quantity to account for the energy losses suffered by the electrons travelling within the solid is the dielectric function.

\subsubsection{Dielectric and energy loss functions}

The energy loss mechanisms due to the electron-electron interaction (other possible energy losses being due to the electron-phonon and, for insulating materials, to polaronic effects) can be modelled via the dielectric formalism  \cite{Ritchie1959}, which relies on the energy- and momentum- dependent macroscopic dielectric function $\bar{\epsilon} ({\bf q},W)$ \cite{Ritchie_PhysRev_1957}. The latter quantity encodes the necessary information on both the single-particle and collective (plasmon) excitations specific of the material. A detailed explanation on how to obtain $\bar{\epsilon} ({\bf q},W)$ from first principles calculations is reported in the appendix.
Within this framework, one relies on the ELF of the material, which is related to the macroscopic dielectric function as follows:
\begin{equation}\label{ELF}
\mathrm{ELF} = \mathrm{Im} \left [\frac{-1}{\bar{\epsilon}(\mathbf{q}, W)} \right ].
\end{equation} 

The ELF is a unique characteristic of the material and, in particular, does not depend on the beam features. It can be obtained by following two routes: one can rely on a fit of the experimental data, typically restricted to the optical regime (long wavelength limit, ${\bf q}\rightarrow 0$), or one can reckon it from first-principles simulations. We stress that the experimental ELF is usually obtained from measured EELS by background subtraction \cite{Tougaard2004}, a procedure that is not free from uncertainty \cite{taioli2010electron,taioli2009mixed,taioli2009surprises} and mostly limited to zero momentum transfer. 

By using first-principles simulations one can determine the ELF of any material without relying on experimental data, also with the inclusion of the momentum transfer dispersion (at the same computational cost of the zero-momentum transfer), which is crucial to determine accurately the IMFP.
It was also shown that the ab initio calculation of the ELF delivers better agreement with experimental measurements of derived quantities, such as the energy loss spectral lineshape as well as the TEY and the dose absorbed by materials \cite{taioli2020relative,pedrielli2021electronic,de2022energy,AZZOLINI2017299}. Thus, we decided to pursue this latter approach.

In particular, we carried out the ab initio calculation of the dielectric linear response function in and out the optical limit (${\bf q} \rightarrow 0)$ using TDDFT implemented in the ELK suite \cite{ELK}. ELK implements an all-electron Full-Potential Linearized Augmented-Plane-Wave approach that is thus not affected by any pseudopotential approximation to describe the ion-electron Coulomb interaction. 
The local spin-density approximation (LSDA) exchange-correlation functional \cite{Perdew1992} has been used for including the spin-orbit (SO) coupling in the spin-polarized calculation of the ground state.

Gold has a face-centred cubic structure belonging to the space group symmetry $Fm-3m$ with a lattice parameter of 4.078 \AA ~\cite{doi:10.1139/p64-213}. A 32 $\times$ 32 $\times$ 32 Monkhorst-Pack $k$-point grid for sampling the Au first Brillouin zone (1BZ) zone and $50$ empty bands in order to obtain converged results up to $\simeq 70$~eV were employed. The ELF were averaged over the three components of the polarization vector of the external electromagnetic field. 

\begin{figure}[htp!]
\centering
\includegraphics[width=1.0\textwidth]{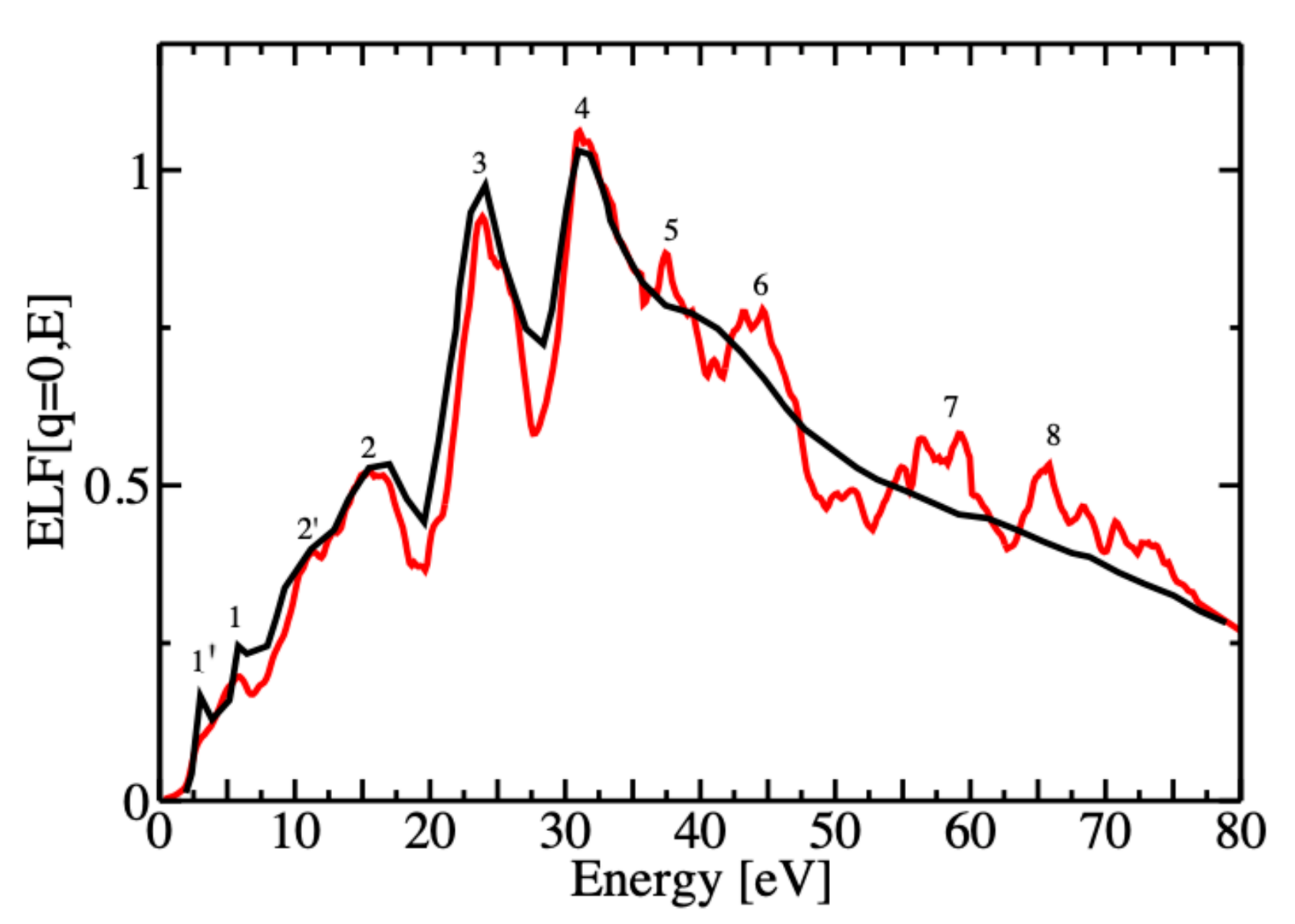}
\caption{Long wavelength limit (${\bf q}\rightarrow 0$) of the ab-initio ELF of Au (red curve) in comparison to experimental optical data from Ref. \cite{werner2009optical} (black curve). 
The LSDA to the exchange-correlation functional has been used to include SO coupling alongside the ALDA approximation to the time-dependent exchange-correlation kernel. }
\label{fig:ELFPOLI1}
\end{figure}

Our ab-initio results (red line) were compared with the available experimental optical data \cite{werner2009optical,RIDZEL2020146824} (black line) in Fig. \ref{fig:ELFPOLI1}, finding an excellent agreement. We notice that the small discrepancies between calculations and experimental measurements of the ELF \cite{werner2009optical,RIDZEL2020146824} in the low energy regime can be rationalized by knowing that the latter are obtained from experimental EELS that include both bulk and surface plasmon excitations. In all our simulations we neglect such surface effects, which can affect the spectral distribution of the emitted electrons, particularly at low energy of the primary beam. 
Comparable results have been obtained in Ref. \cite{PhysRevB.102.035156}, where the ELF of Au was obtained by using a different approach based on Liouville-Lanczos method adopting scalar relativistic ultra-soft pseudopotentials to model the ion-valence electron interaction and the Lanczos recursion method that avoids computationally-expensive summation over empty bands \cite{PhysRevB.88.064301,TIMROV2015460}. In Tab. \ref{tab:interpret} we report the peak energy positions together with their theoretical interpretation in comparison to the analysis reported in Refs. \cite{PhysRevB.102.035156,GURTUBAY2001123}. While our findings substantially agree with those in Ref. \cite{PhysRevB.102.035156}, we are tempted to assign the peaks at 23.74 and 31.35 eV (peak 3,4 in Fig. \ref{fig:ELFPOLI1}) to mixed collective excitations, owing to an almost zero real part of the dielectric function along with a small while not negligible imaginary part accounting for their widths (as found in Refs. \cite{GURTUBAY2001123,PhysRevB.88.195124}, where a plot of Re$(\epsilon)$ and Im$(\epsilon)$ is reported). 
We remind that the ELF in Eq. \ref{ELF} can be rewritten as
\begin{equation}
\mathrm{ELF}=-\mathrm{Im}(\bar{\epsilon}^{~-1})=\frac{\mathrm{Im}(\bar{\epsilon})}{\mathrm{Re}(\bar{\epsilon})^2+\mathrm{Im}(\bar{\epsilon})^2}
\end{equation}
which implies that at this frequency the plasmon resonance (determined by the fulfillment of the condition Re$(\bar{\epsilon})\simeq 0$) is influenced by the interband transitions therein (Im$(\bar{\epsilon})\neq 0$).

\begin{table}[h!]
\caption{Optical ELF (${\bf q} \rightarrow 0$) of bulk Au. First column: peak labels according to Fig. \ref{fig:ELFPOLI1}. The second ($\omega^{a}$) and third ($\omega^{b}$) columns report the peak energy position in the ELF as by Refs. \cite{PhysRevB.102.035156} and \cite{GURTUBAY2001123}, respectively. 
        In the fourth column we report the values obtained in this work with the number of digits representative of the numerical accuracy, while in the fifth column (Exp.) the EELS experimental data from \cite{werner2009optical}. In the last column (Origin) the theoretical interpretation of each energy loss peak is outlined.
         IT stands for interband transition, while ME means mixed excitation.}
         \label{tab:interpret}
    \begin{indented}
         \lineup
        \item[]\begin{tabular}{@{}c c c c c c}
        \br 
        \multicolumn{1}{c}{peak label} &
        \multicolumn{1}{c}{$\omega^{a}$ (eV)}  &
        \multicolumn{1}{c}{$\omega^{b}$ (eV)} & 
        \multicolumn{1}{c}{This work} &
        \multicolumn{1}{c}{Exp.} &
        \multicolumn{1}{c}{Origin}  \\
        \mr
        1'   &    2.2  &   &  2.66 &  2.5  & ME   \\ 
        1   &   5.1  &  5.5 &  5.94 &  5.9   & 5d plasmon    \\ 
        2'   &  10.2   & 11.5 &  11.49 & 11.9  & Mainly-6s plasmon     \\ 
        2   &   15.5  &	 18 &    15.67 & 15.8  &  IT     	 \\
        3   &   23.8  &	25 &   23.74 & 23.6 &  ME  \\
        4   &   30.8  &	35 &    31.35 & 31.5 &  ME          	 \\
        5  &    36.9 & 40.5	&    37.5 & 39.5  & IT              \\
        6   &   43.5. &	43.5 &    44  & 44 & IT      \\
         7  &   &	&    59.2  &  &	IT   \\
          8   &  & 	&   65.62  & &  IT  \\
        \br 
        \end{tabular}
    \end{indented}
\end{table}

\begin{figure}[htp!]
\centering
\includegraphics[width=0.8\textwidth]{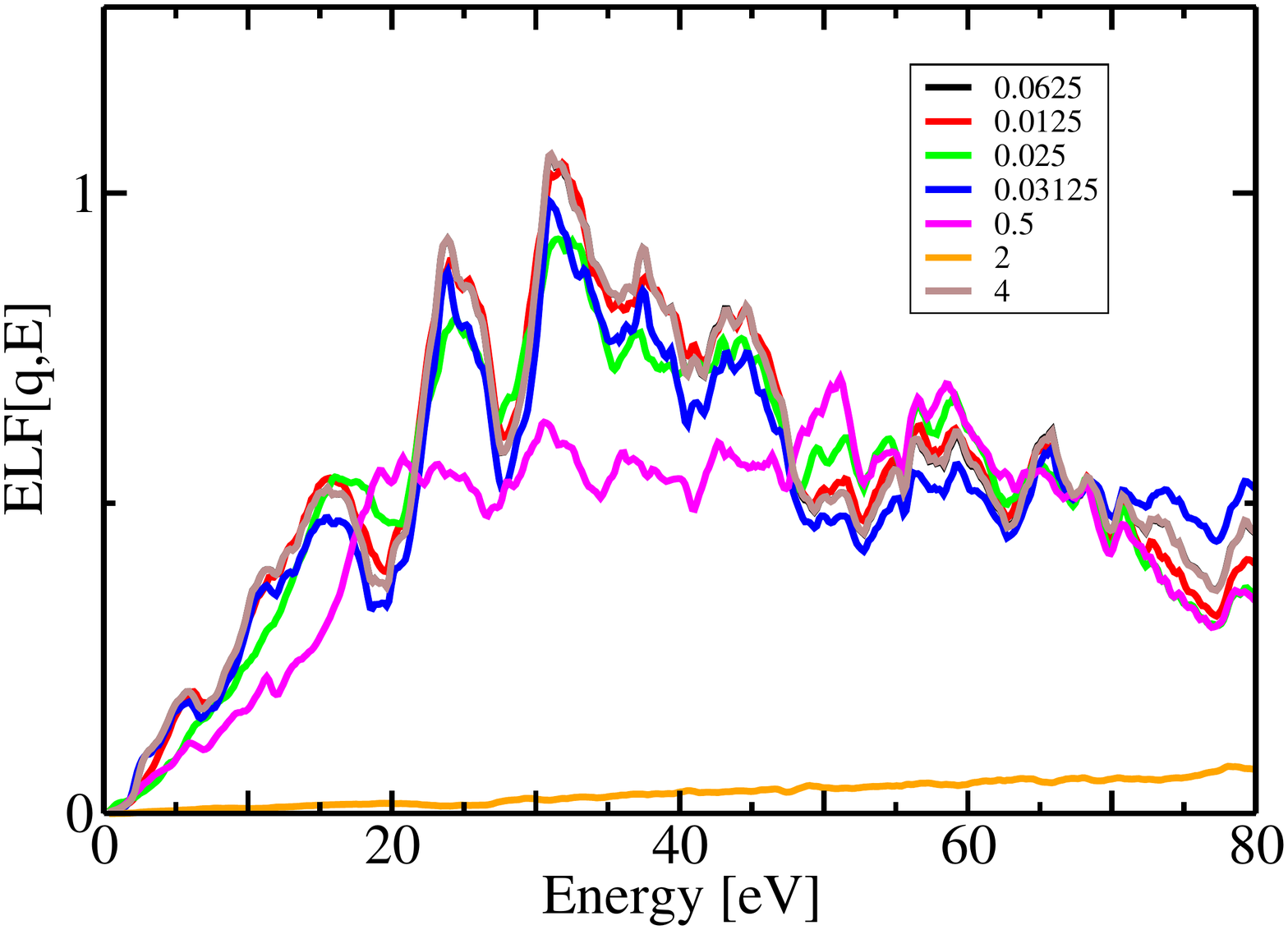}
\includegraphics[width=0.8\textwidth]{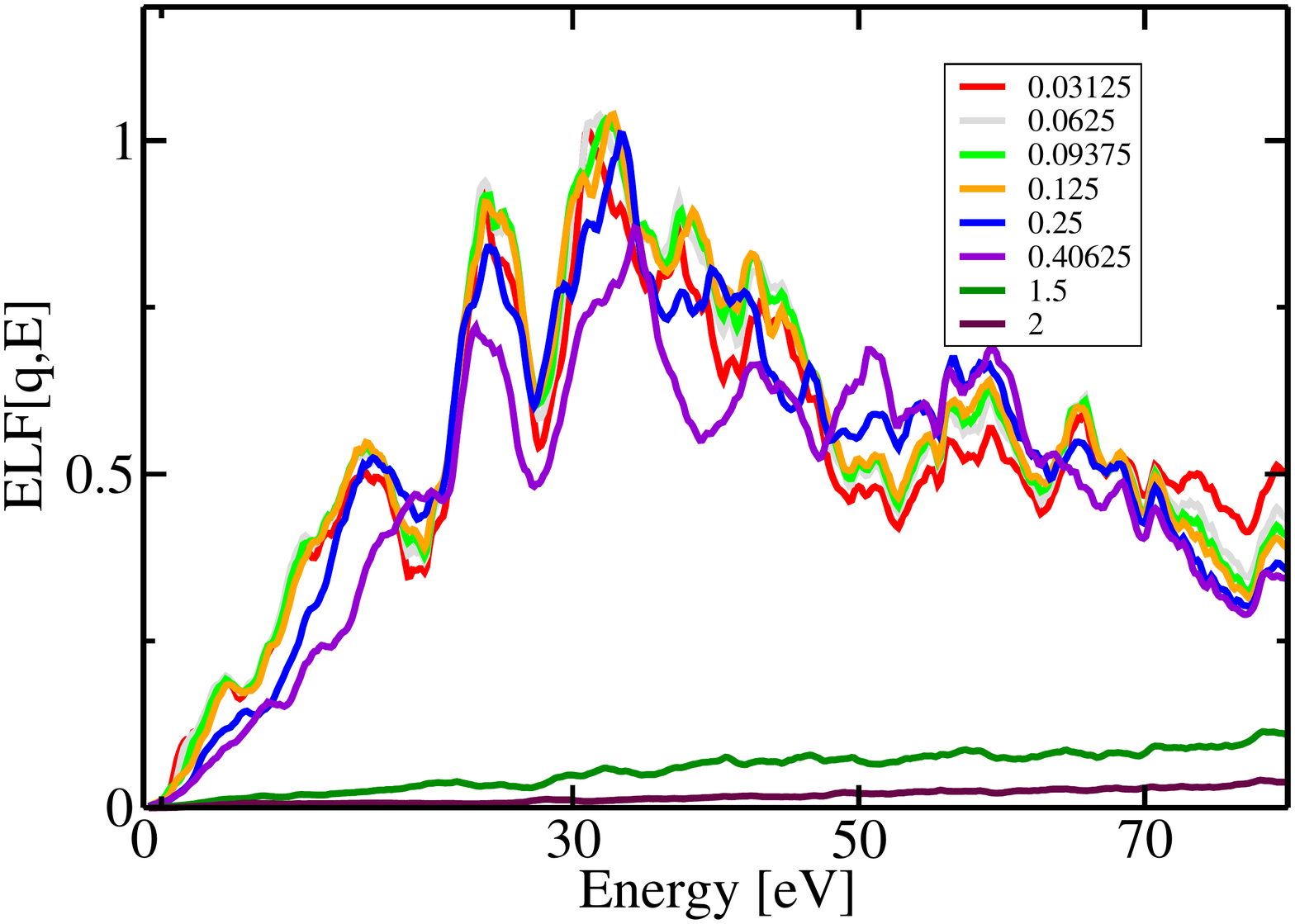}
\caption{Top panel: ELF of Au at finite momentum transfers along the (1,1,1) direction. The calculated $q$-vectors are reported in lattice coordinates in the legend (e.g. 0.0625 means the $q$-vector (0.0625, 0.0625, 0.0625)).
Bottom panel: ELF of Au at finite momentum transfers along the (1,1,0) direction. 
The calculated $q$-vectors are reported in lattice coordinates in the legend (e.g. 0.0625 means the $q$-vector (0.0625, 0.0625, 0)).
The LSDA to the exchange-correlation functional has been used to include SO coupling alongside the ALDA approximation to the time-dependent exchange-correlation kernel.}
\label{fig:ELFPOLI3}
\end{figure}

We stress that the well-resolved double-peak shape of the ELF is due to the presence of localized $d-$bands, which are typical of other transition metals. The use of the LDA approximation to describe the $d-$bands of Au may lead to 0.5 eV redshift of the interband transitions in this energy region (see e.g. \cite{PhysRevB.88.195124}), an effect that can be partially cured by introducing many-body effects via ab initio many-body perturbation theory, whereby a $\simeq$ 0.4 to $\simeq$0.8 eV opening of the $5d-6sp$ interband gap may reduce the discrepancy with the experiments \cite{PhysRevB.86.125125}. We also notice in Fig. \ref{fig:ELFPOLI3} that the double-peak shows little dispersion and lower intensity at finite momentum transfers along the two different symmetry directions (1,1,1) (top panel of Fig. \ref{fig:ELFPOLI3}) and (1,1,0) (bottom panel of Fig. \ref{fig:ELFPOLI3}). 
This almost dispersionless characteristic suggests that their nature resembles interband electronic transitions rather than collective excitations. This seems also the case of the shoulder at 2.66 eV (1' in Fig. \ref{fig:ELFPOLI1}) and at 15.67 eV (peak 2 in Fig. \ref{fig:ELFPOLI1}), which show a very weak dispersion with increasing momentum transfer. We stress that in EELS experimental data the low-energy shoulder overlaps with the surface plasmon contribution \cite{Yoshikawa}. The peaks at 5.94 and 11.49 eV (peaks 1 and 2' in Fig. \ref{fig:ELFPOLI1}), which have been identified as 5$d$ and 6$s$ plasmons, display a sizeable momentum dispersion (more evident in the bottom panel of Fig. \ref{fig:ELFPOLI3}), which reinforces their interpretation as collective charge excitations. In fact, besides the double-peak structure, the localized $d-$bands play another important role in the dielectric response function of Au for damping the $5d$ plasmon energy from that expected by the free-electron plasma frequency model ($\simeq$9 eV) to 5.94 eV (peak 1 in Fig. \ref{fig:ELFPOLI1}). These two well-resolved plasmon peaks show that $5d$ and $6s$ valence electrons in Au behave as separate electron gases. In general, at very high momentum transfers (see orange and purple lines in the top and bottom panels of Fig. \ref{fig:ELFPOLI3}) the spectral lineshape is characterised by a flat behaviour; in this regime the transitions are single-particle in nature and dominated by the kinetic energy. Peaks at higher energies (5,6,7, and 8 in Fig. \ref{fig:ELFPOLI1}) are also weakly dispersing, which indicates their nature of interband transitions.

\subsubsection{Fit of the ELF by classical Drude oscillators}

To run our transport MC scheme for modelling the energy loss and secondary emission spectra of Au, we need to determine also the inelastic scattering cross section. The latter can be obtained by knowing the dependence of the ELF -- the so-called Bethe surface -- over the entire spectrum of excitation energies $W$ and momentum transfers ${\mathbf{q}}$ \cite{AZZOLINI2017299,azzolini2018anisotropic,taioli2020relative,pedrielli2021electronic,azzolini2020comparison}.
The first-principles calculations of the ELF over a large range of energies is prohibitive due to the enormous number of electronic transitions to the excited states that must be included in the model. 
Moreover, the richest structure can be found in the low-lying part of the  excited spectrum, in an energy range of 80 eV from the Fermi level, the high energy region being characterised typically by a few core level vertical transitions and a pre-continuum broad lineshape. Thus, the extension of the ELF along the excitation energy axis has been carried out by matching the ab-initio results with the experimental NIST X-ray atomic data set up to 30 keV \cite{NIST} using $E_{5p1/2} = 74.2 \ $ eV \cite{thompson2001x} as threshold energy of the semi-core transitions for Au.

\begin{table}[h!]
        \caption{D--L parameters of the optical ELF (${\bf q} \rightarrow 0$)}
         \label{tab:fitELF1}
    \begin{indented}
         \lineup
        \item[]\begin{tabular}{@{}r r r r}
        \br 
        \multicolumn{1}{c|}{\textbf{$i$}} &
        \multicolumn{1}{c|}{$A_i$ (eV$^2$)}  &
        \multicolumn{1}{c|}{$\gamma_i$ (eV)}    &
        \multicolumn{1}{c}{$W_i$ (eV)} \\
        \mr
       
        1   &   1.969      &   3.463  & 5.382   \\ 
        2   &   17.508      & 	6.121  &	  11.644 \\ 
         3   &   26.815     &   4.948  & 	 15.909 \\ 
         4   &   106.586 	&   5.784  & 24.449	 \\
        5   &   43.843 	&   3.086  & 	 30.993 \\
        6   &   98.402 	&   5.881 & 33.355	 \\
        7   &   90.344 	&   5.670 & 37.938	 \\
        8   &   177.047 	&   7.852 & 44.489	 \\
        9   &   551.321 	&   18.326 & 59.799	 \\
         10   &  29.773   	&   3.667 & 68.164 \\
        11   &  128.517   	&   7.449 & 74.020	 \\
        12   &  65.4861   	&   6.292 & 87.151  \\
        13   &    46.896 &   7.192 &  95.522 \\
        14   &  2500.0   &   350.0 & 330.0 \\
        15 & 2000.0  & 2000.0 & 2500.0 \\
         16 &  200.0 	&   5000.0 & 14000.0 \\
        \br 
        \end{tabular}
    \end{indented}
\end{table}

\begin{figure}[h!]
\centering
\includegraphics[width=1.0\textwidth]{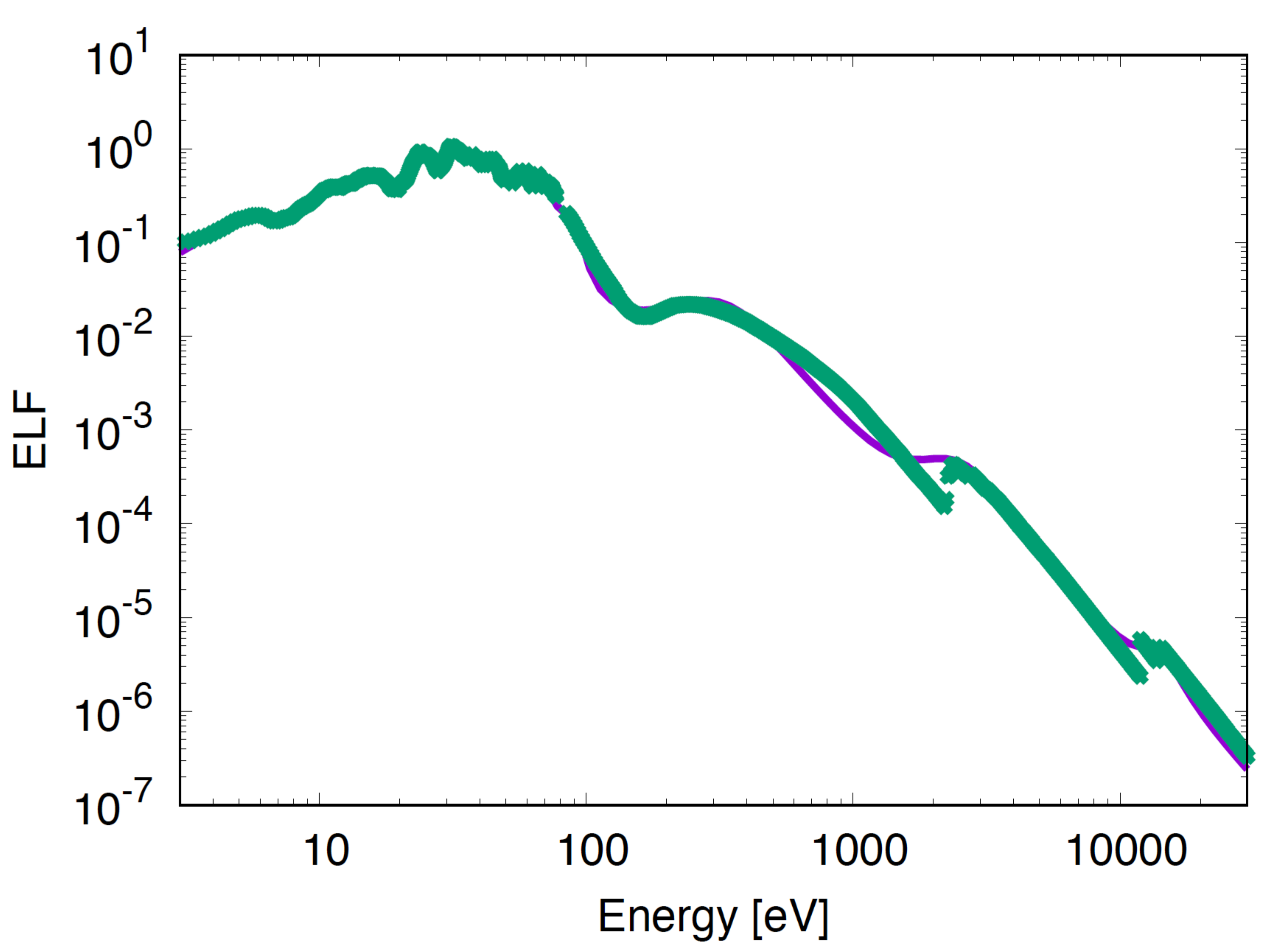}
\caption{Ab-initio ELF (green line) extended to high excitation energies \cite{NIST} alongside the D-L fit (purple line). The fit parameters are reported in Tab. \ref{tab:fitELF1}.}
\label{fig:ELFPOLI2}
\end{figure}

Finally, the total ELF in the optical limit was fitted by using generalized classical Drude-Lorentz (D--L) oscillators as follows:
\begin{equation}\label{Im}
{\mathrm{Im}}\Big[\frac{-1}{\bar{\epsilon}({\bf q}=0),W)}\Big]_{\rm}=\sum_{i} \frac{A_i \gamma_i W}{(W_i^2-W^2)^2+(\gamma_i  W)^2} 
\end{equation}
where $A_i$, $W_i$, and $\gamma_i$ are fitting parameters, which physically represent the strength, energy and linewidth of the electronic transitions, respectively. In Tab. \ref{tab:fitELF1} the optimal fitting parameters of the D--L functions used to fit the optical ELF (zero momentum transfer) are reported. In the fitting procedure we checked that the $f$-sum rule is satisfied. In Fig. \ref{fig:ELFPOLI2} we show the fit (green line) of the ab-initio ELF extended up to 10 keV with the NIST data \cite{NIST}. We notice (see Fig. \ref{fig:ELFPOLI1}) that the TDDFT ELF data have a significantly higher resolution than the optical data, as detailed optical measurements in this regime can be difficult. This results in the need of including several Drude oscillators to reproduce the ab-initio data.

Finally, to model the extension of the ELF to finite momentum transfers (${\bf q}\neq 0$) along the entire energy axis, we found that the momentum dispersion of the electronic excitations obtained from ab-initio calculations can be accounted for by introducing a dispersion law to the characteristic energies of the oscillators $W_i(q)$ as follows \cite{Ritchie1977}:
\begin{equation}\label{disp}
W_i(q) = \sqrt{{W_i}^2+(12/5) \cdot E\mathrm{_f} \cdot  q^2/2+q^4/4}
\end{equation}
where $E_{\mathrm{f}}$ is the Fermi energy. 

\subsubsection{The inelastic mean free path of electrons in Au}

\begin{figure}[h!]
    \centering
    \includegraphics[width = 1.0\textwidth]{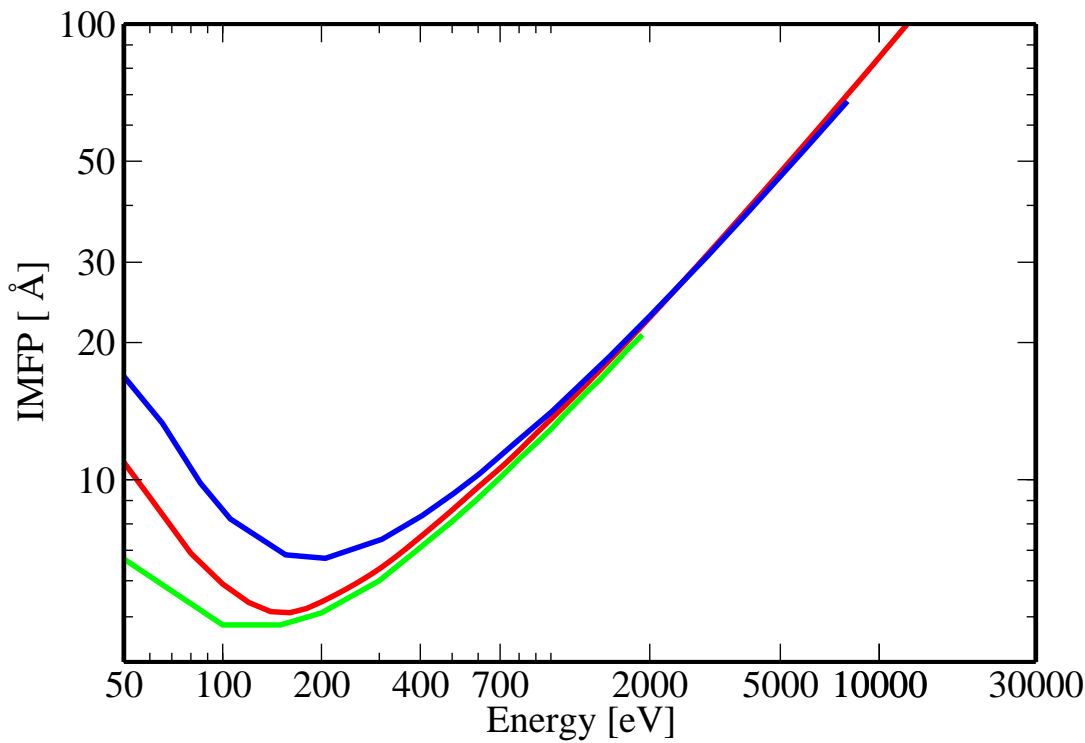}
    \caption{Inelastic mean free path (red line) of Au as a function of the primary beam kinetic energy, in comparison to Tanuma et al. (green curve) \cite{https://doi.org/10.1002/sia.740171304} and Ashley et al. (blue curve) \cite{ASHLEY1990323} 
    }
    \label{fig:imfp}
\end{figure}

The knowledge of the ELF as a function of both transferred momentum and energy allows one to calculate the differential inverse inelastic mean free path, using the following relation:
\begin{equation}\label{diimfp}
    \frac{d \lambda_\mathrm{inel}^{-1}}{dW} = \frac{1}{\pi a_0 E}\int_{q_-}^{q_+} \frac{dq}{q}[1+f_{\mathrm{exc}}(q)] \mathrm{Im}\left[-\frac{1}{\bar{\epsilon}(q, W)}\right]
\end{equation}
where $a_0$ is the Bohr radius, $m$ is the electron mass, and the integration limits are $q_\pm = \sqrt{2mE} \pm \sqrt{2m(E-W)}$.
In Eq. \ref{diimfp} we introduce the exchange interaction between the impinging and target electrons according to Born-Ochkur \cite{Ochkur}, which reads:
\begin{equation}\label{bokur}
f_{\mathrm{exc}}(q)=\left( \frac{\hbar q}{mv}^4\right) -\left( \frac{\hbar q}{mv}^2\right)
\end{equation}
where $v=\sqrt{2E/m}$ is the electron velocity. 
The IMFP can be obtained by integrating the equation \ref{diimfp} in the interval of energy losses (0, $E/2$) (assuming that the secondary electrons are those emerging with lower kinetic energy). Our results are shown in Fig. \ref{fig:imfp} (red line) and compared with calculations presented by Tanuma et al. (green line) \cite{https://doi.org/10.1002/sia.740171304} and Ashley et al. (blue line) \cite{ASHLEY1990323}.

As a result of inelastic scattering, electrons lose a fraction $\overline{W}$ of their kinetic energy. In the SEED statistical model, such energy loss is determined by equating a 
random number $\mu_3$ uniformly distributed in the interval [0,1] with the relevant inelastic cumulative probability: 
\begin{equation}
    P_\mathrm{inel}(E,\overline{W}) = \lambda_\mathrm{inel}(E) \int_0^{\overline{W}} \frac{d \lambda_\mathrm{inel}^{-1}(E,W)}{dW} dW
\end{equation}

The value of $\overline{W}$ for which $P_\mathrm{inel}=\mu_3$ is regarded as the energy loss due to a single inelastic interaction. The electron kinetic energy is then decreased by this value. 
However, if the energy loss is larger than the first ionization energy $B$, an ionization occurs: a secondary electron will be generated which starts traveling within the solid target with kinetic energy $E= W - B$ and may undergo, in turn, elastic and inelastic scattering processes. Here we assume that in gold, work function and ionization energy $B$ are the same (4.7 eV).
Having access to the elastic and inelastic scattering cross sections, our SEED algorithm works as follows: we extract 
a random number $\mu$ uniformly distributed in the interval [0,1], and we use such random number to determine the step length that each electron travels within the solid target $\Delta s=-\lambda \ln(\mu)$, where $\lambda$ is total mean free path. $\lambda$ includes all the scattering processes and can be assessed from $\lambda^{-1}=\lambda^{-1}_\mathrm{el}+\lambda^{-1}_\mathrm{inel}$, where $\lambda^{-1}_\mathrm{el}$ and $\lambda^{-1}_\mathrm{inel}$ are the inverse elastic and inelastic mean free paths, respectively.
Finally, the SEED method proceeds by selecting either an elastic or inelastic scattering event by comparing a random number $\mu_4$, uniformly distributed in the range $[0,1]$, with the relevant scattering probability. In particular, if $\mu_4< p_{el}=\lambda_{el}^{-1}/\lambda^{-1}$ the scattering is elastic; otherwise the electron faces an inelastic event. 

\subsection{Energy loss spectrum and secondary electron total yield of Au}

\begin{figure}[h!]
    \centering
    \includegraphics[width = 1.0\textwidth]{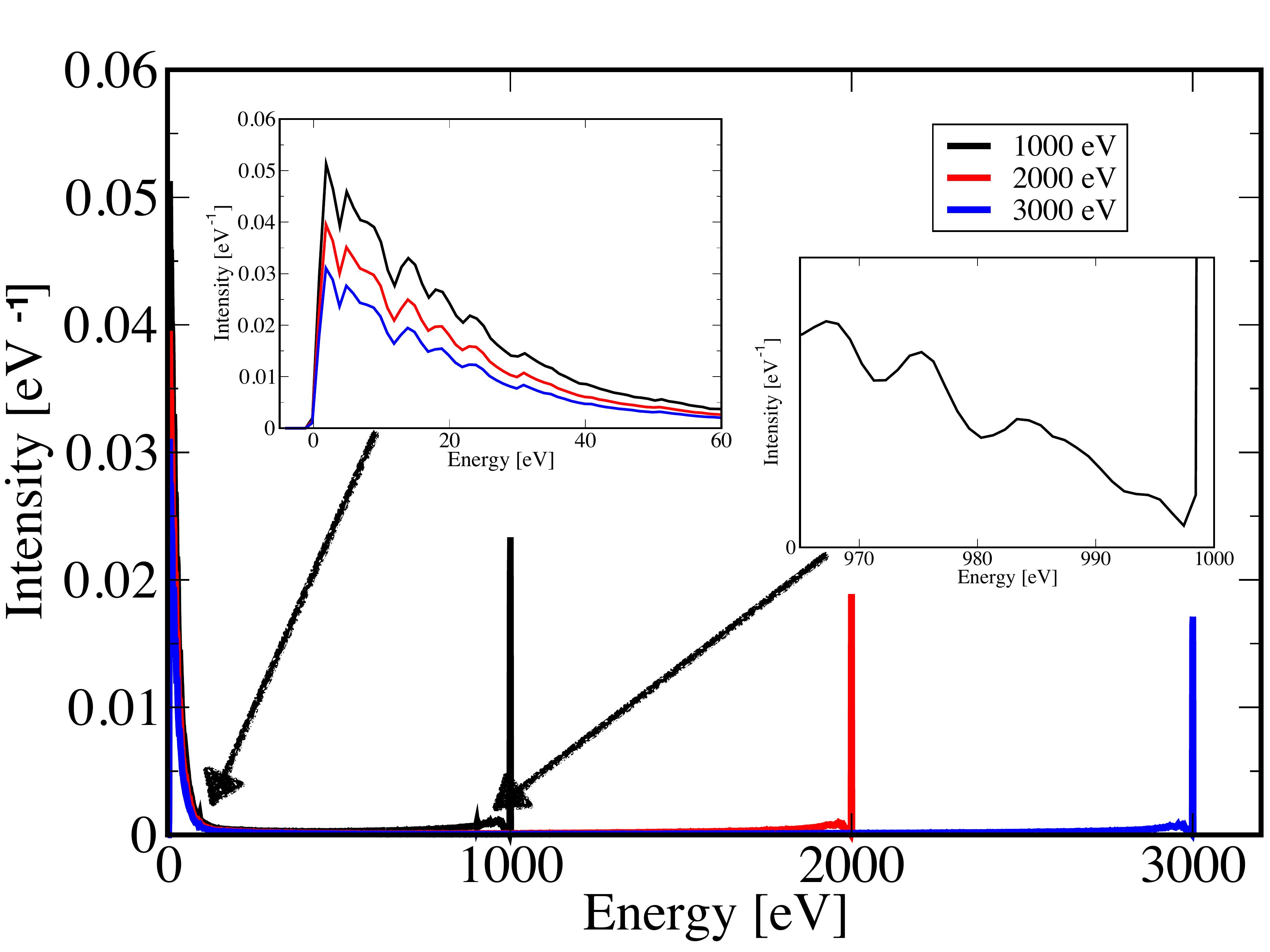}
    \caption{Monte Carlo simulation of the energy-loss spectrum of electrons emerging from the Au sample for different primary electron beam kinetic energies: black line (1000 eV), red line (2000 eV), blue line (3000 eV). Left and right insets report the secondary electron distribution (for all primary beam kinetic energies) and the plasmon peaks (for 1000 eV only), respectively.}
    \label{fig:eloss}
\end{figure}

The elastic and inelastic cross sections of Au allow SEED to simulate a simple backscattering experiment of an electron beam on a thick gold target, and to calculate the energy spectrum of the emerging electrons shown in Fig. \ref{fig:eloss}.
In our SEED simulations we include the most important mechanisms of energy-loss, such as the single-electron excitation from valence to conduction bands, plasmon generation, and the elastic scattering with the ions of Au. These energy loss mechanisms 
are reproduced as several well-resolved peaks in the experimental energy loss spectrum, which collects the electrons emerging at a certain angle from the Au specimen as a function of their kinetic energies.
A fraction of the electrons of the primary beam can be backscattered with the same energy as the primary beam. These electrons form the so-called elastic -- or zero-loss -- peak. 

In Fig. \ref{fig:eloss} the elastic peak can be found at the extreme right of the relevant spectrum as a narrow peak centered at the kinetic energy of the primary beam (1000 (black line), 2000 (red line), and 3000 (blue line) eV, respectively). The half width at half maximum is set to 0.4 eV to account for the typical finite resolution of an electron spectrometer and the beam's intrinsic width.

We now discuss the spectral lineshape of Fig. \ref{fig:eloss}: we use the energy loss as a descriptor because of its invariant value with respect to the primary beam kinetic energy.
The shoulder at $\simeq$ 2.3 eV below the elastic peak (mixed excitation), which  cannot be attributed to surface plasmon losses not included in our calculation, are not clearly resolved, owing to the peak broadening. Also the very low energy loss due to the electron-phonon scattering (fraction of eV) are hidden behind the elastic peak.  Moving towards larger energy loss, the spectrum presents the plasmon peak (see the inset at the right in Fig. \ref{fig:eloss}), which is due to the electrons that have suffered a single inelastic scattering within the medium and were able to excite a plasmon. The bulk plasmon peaks of Au can be found at $\approx$ 5.77 and 11.6 eV on the left of the elastic peak, slightly shifted with the respect to the maxima of the ELF (5.94 and 11.49 eV, respectively, see also Table \ref{tab:interpret}). Of course, multiple electron-plasmon scattering and excitation may occur; however, they have very low probability and, thus, intensity (they are scarcely visible in the spectra of Fig. \ref{fig:eloss}). Finally, the origin of the series of clearly resolved spectral features (see the inset at the right in Fig. \ref{fig:eloss}) found in the region 15-65 eV from the elastic peak at $\approx$ 15.7, 24.7, 31.7, and 44 eV can be attributed to $5d,4f$ $\rightarrow$ 6s interband electronic transitions, with the exception of the intermediate peaks at 24.7 and 31.7 eV, which show also plasmon-like characteristics (Re$(\epsilon) \simeq 0$) affected by the concurrent contribution of interband and intraband transitions occurring around the same energy (Im$(\epsilon)$ small but $\neq 0$).

The region of the spectrum characterised by a pronounced broad peak below 50 eV (see the inset on the left side of Fig. \ref{fig:eloss} zooming this part of the energy loss spectrum), which is called the secondary electron peak, collects those electrons that have been extracted by electron-electron collisions and emerge from the target surface after having lost most of their energy via several inelastic scattering events. 
Owing to the number of well-resolved peaks in the ELF, it is not surprising that our MC simulations of the secondary electron peak (see the left inset of Fig. \ref{fig:eloss}) find a wiggling rather than a smooth behaviour in that energy region. Secondary electrons are generated upon inelastic scattering events when the primary beam electrons lose a kinetic energy corresponding to the ELF peaks. Furthermore, secondary electrons can lose in turn their kinetic energy before leaving the surface of the sample according to the ELF. These wiggling features are also present when we use a very high number of trajectories (in excess of 10$^9$), and thus cannot be attributed to statistical noise. 

Using the energy-straggling strategy implemented in our MC code \cite{Dapor_book_blu}, we focused also on the quantitative modelling of the secondary electron emission yield from Au. In particular, we reproduce by SEED the TEY, as the sum of both backscattered and secondary electrons for primary beam kinetic energies in the range 0-10000 eV, that is the integral of each of the energy loss spectra at each primary kinetic energies.  

If the impinging particles have a kinetic energy above the ionization threshold of the atomic constituents, secondary electrons can be ejected upon inelastic scattering events and, after suffering in turn a number of elastic and inelastic interactions, may 
eventually escape from the target surface, provided that the electron overcome the energy barrier at the solid-vacuum interface (work function). The angular emission of the secondary electrons is reckoned by adopting the classical binary collision model \cite{de2022energy}, taking into account momentum and energy conservation. 

The emission condition at the solid-vacuum interface reads: 
\begin{equation}
E \cos^2 \theta \ge \Phi,
\label{eq:emissione}
\end{equation}
where $\theta$ describes the angle formed by the normal to the sample surface and the electron trajectory intercept, and $E$ is the residual electron kinetic energy. 
The transmission coefficient $T$ can be obtained by modelling the electron emission as a problem of a particle scattered by a step potential barrier \cite{Dapor_book_blu}: 
\begin{equation}
    T = \frac{4\sqrt{1 - \Phi/(E\cos^2 \theta)}}{\left[1 + \sqrt{1 - \Phi/(E\cos^2 \theta)} \right]^2}.
\end{equation}

The electron emission from the target surface occurs, according to our statistical scheme, by
comparing $T$ with a random number $\mu_5$ uniformly distributed in the interval [0,1]:  if $\mu_5 < T$ the electron emerges from the surface and its energy is decreased by an amount equal to the work function $\Phi$ of the material (4.7 eV for Au), otherwise it is scattered back to continue its path within the solid target. 

\begin{figure}[h!]
\centering
\includegraphics[width=1.0\textwidth]{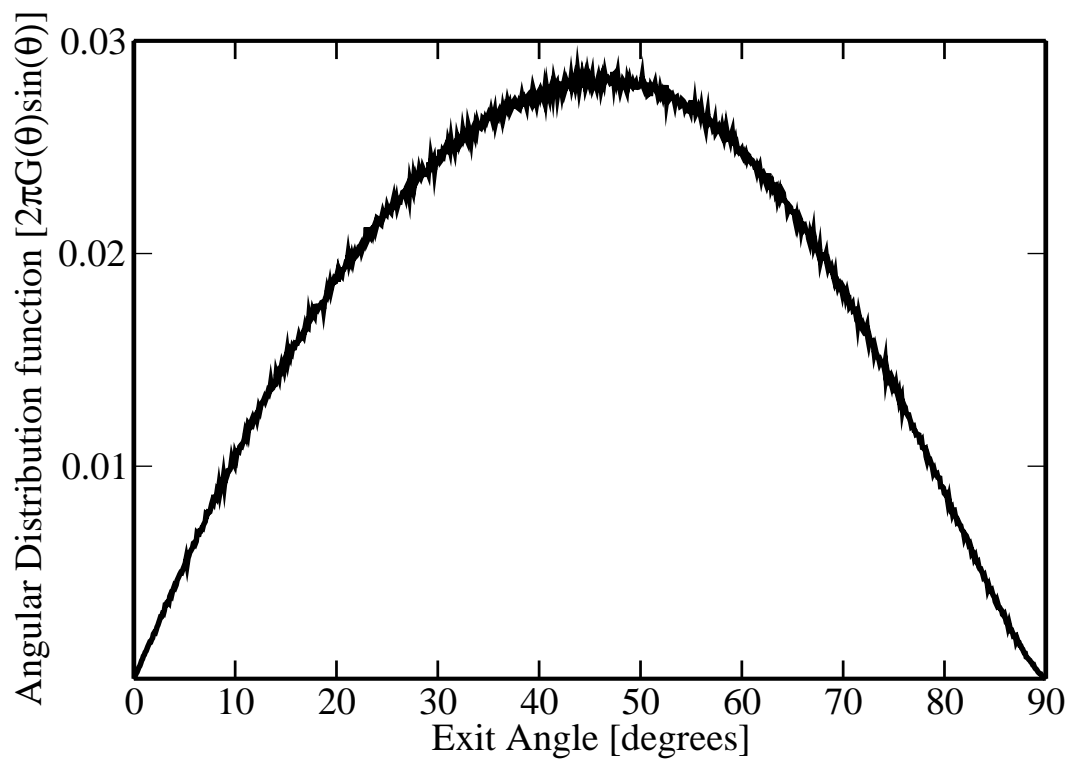}
\caption{Cosine angular distribution $G(\theta)$ of SE escaping from the Au surface for 1 keV primary electron beam.}
\label{fig:angdist}
\end{figure}

We notice that the value of $\Phi$ has a paramount importance in determining the TEY and in describing the electron emission process:
in particular, the higher is the work function (the barrier to overcome) the smaller is the electron yield. 
In Fig. \ref{fig:angdist} we show the angular distribution pattern of the secondary and backscattered electrons emitted from the Au surface upon 1 keV electron beam impinging on the sample. In particular, the angular distribution follows closely a cosine function with a maximum $\approx$ {45\textdegree}  and is almost inversion symmetric around that value. Notice that such distribution is produced by normalizing to the value of the yield (1.5) we have obtained by using SEED at 1 keV.

\begin{figure}[h!]
\centering
\includegraphics[width=1.0\textwidth]{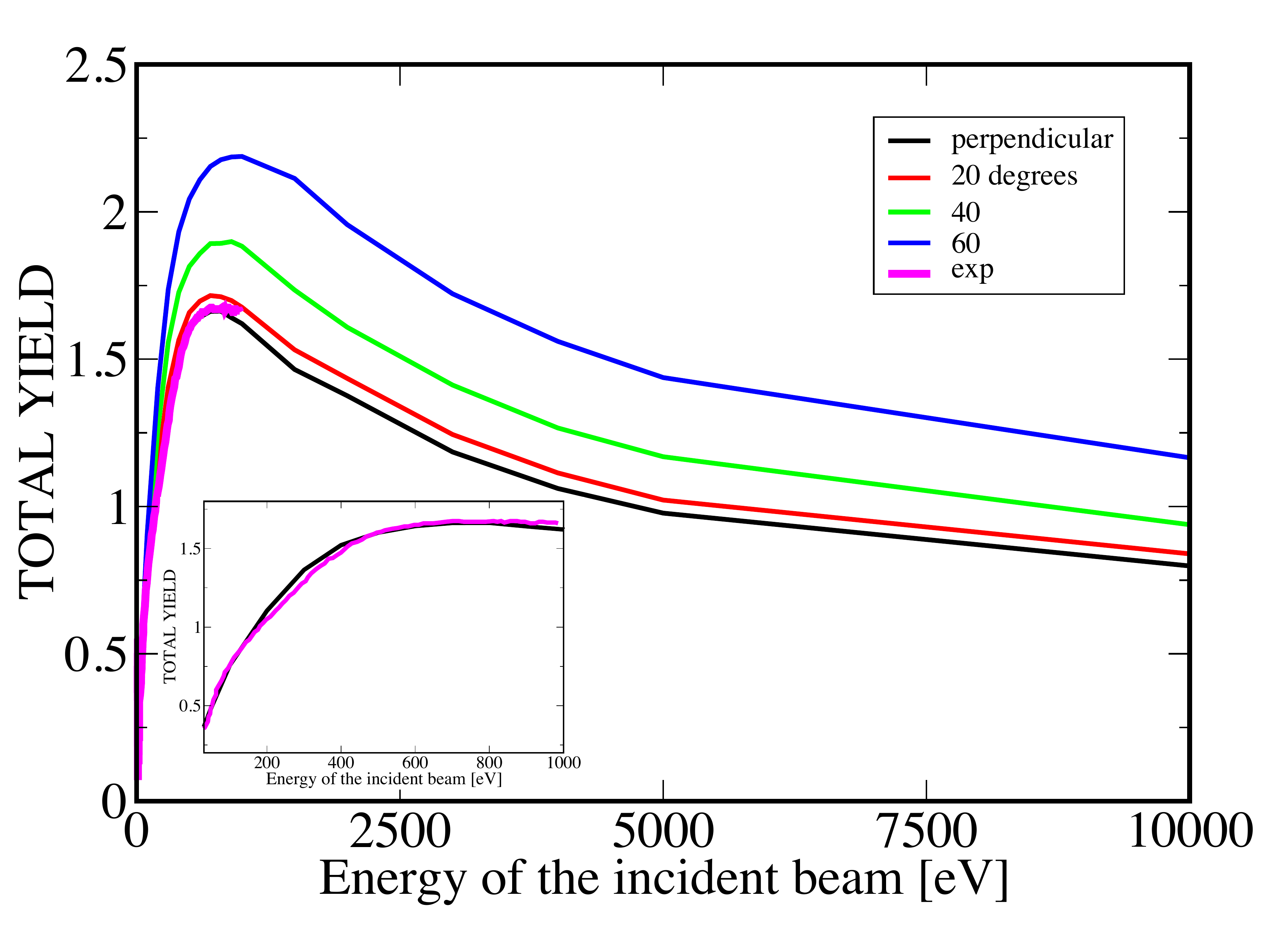}
\caption{Comparison between our SEED calculations of TEY from an Au sample as a function of the primary beam initial kinetic energy and the experimental data (magenta line) \cite{gonzalez2017secondary} for $\mathrm{Ar}^+$ sputter cleaned Au in situ. Calculations were carried out for different incident angles of the electron beam; black curve: normal incidence; red, green, and blue lines report the TEY for incident angles of 20$^\circ$, 40$^\circ$, and 60$^\circ$ with respect to the surface target normal, respectively. In the inset we zoom around the energy region where experimental data are available.}
\label{fig:yield}
\end{figure}

In Fig. \ref{fig:yield} we report the TEY as a function of the initial kinetic energy of the primary beam in comparison to experimental data (magenta line) for $\mathrm{Ar}^+$ sputter cleaned Au in situ \cite{gonzalez2017secondary}. In particular, the black line shows the yield for normal incidence of the primary beam with respect to the target surface, while the red, green, and blue lines report the yield for incident angles of 20$^\circ$, 40$^\circ$, and 60$^\circ$ with respect to the target surface normal, respectively. We notice that the agreement between the black curve and the experimental results, recorded at the same geometry, is remarkable (see the inset of Fig. \ref{fig:yield} in which we zoom around the energy region of the available experimental data). 

Furthermore, we notice that the yields at all incident angles display an increasing behaviour for increasing primary beam kinetic energy until a maximum is reached. At very high energy ($>30$ keV, not reported in Fig. \ref{fig:yield}) the yield reaches an asymptotic value of $\approx$ 0.5, which is the known value of the backscattering coefficient of Au. 
This behaviour can be easily rationalized by  
assuming that both the number and the average depth at which the secondary electrons are produced increases with the primary beam kinetic energy. Thus, at low primary energy a relatively small number of secondary electron is produced close to the surface. At increasing primary energy, the average depth of secondary emission increases (as so does the number of secondary electrons generated, owing to the higher primary kinetic energy) to the point that it becomes so deep that only a small number of secondary electrons is eventually able to escape from the target. Our SEED approach is able to reproduce this behaviour, with the maximum monotonically blueshifting and becoming more intense with increasing incident angles with respect to the surface normal. Indeed, for grazing incident angles of the incoming electron beam, the number of escaping secondary electrons increases.

{ In the end, we note that surface contamination of the gold sample due to air exposure prior to the flights can have a dramatic impact on its properties, in particular on the work function $\Phi$ \cite{LPF_Dan},\cite{turetta:hal-03240126},\cite{YANG201688},\cite{wass2019}.  In this respect, surface treatments, bake out cycles, and the local pressure environment can have an important and not easily predicted influence on the actual value of the work function. The most frequent contaminants on gold samples exposed to air are carbon and oxygen (see Ref. \cite{turetta:hal-03240126} and \cite{YANG201688}).
According to Turetta et al., their presence leads to a fluctuation of $\Phi$ in air from a value of $5.25$ eV for clean Au(111) to $4.75$ eV as a result of the exposure to laboratory environment for a short time (few minutes), and back to $5.25$ eV after a relatively long period (few months) (see Fig. 1 a),b) of  Ref. \cite{turetta:hal-03240126}). 
Perhaps the most relevant case for LISA hardware is the in-flight experience from LPF, where photoelectric currents were measured under UV illumination -- at $254$nm, $4.88$eV -- by the charge management system \cite{LPF_Dan}.  The effective data for photoelectric yield as a function of TM potential were compatible with work functions from the TM and EH gold coatings in the range of $3.9$ to $4.5$ eV.  These surfaces had, after exposure to air during assembly, been baked to $115$°C in vacuum before final sealing and storage on ground for roughly $1$ year before launch, followed by venting to deep space in orbit. Work functions in the same range were observed by Wass et al. for gold samples under vacuum condition after limited air exposure \cite{wass2019}, although one sample, out of five examined, exhibited a higher ($>4.89$ eV) value after more than $2$ years in an uncontrolled vacuum environment. 

The TEY calculations were carried out for a value of $\Phi$ equal to $4.75$ eV, achieved within the first $60$ minutes of air exposure according to Turetta's observations, which represents an average value between the clean Au(111) and the LISA expectations \cite{wass2019}. We note that the work function is an input parameter of our simulation; by repeating the simulations for $\Phi=5.25$ eV, we found a decrease of $\approx~10\%$ of the TEY in the spectrum of primary beam kinetic energies investigated in this work. Decreasing the work function would indeed result in an increment of the TEY.}

The SEED code uses a CPU time that is almost proportional to both the number of trajectories included in the simulation and to the primary beam kinetic energy, and is also dependent on the work function, whose value affects the number of emitted secondary electrons. Typical CPU times for generating the electron spectrum using $10^7$ trajectories, with a work function equal to $5.25$ eV, for a primary beam kinetic energy of $1$ keV, are $\approx 2000$ sec for a modern CPU equipped with $10$ cores. We note that this time performance exceeds by far the GEANT4 one.

\section{Perspectives for the study of the LISA TM charging}
\label{s3} 

The results of the SEED simulation reported in Sect. \ref{s2}  contain new relevant information about the charging process of the LISA TMs. The spectra in Figure \ref{fig:eloss} indicate that energy of the backscattered electrons has an important component below 50 eV due to the secondary electrons: for $1$ keV beam, $\approx 25\%$ of the emerging electrons are in the $0-10$ eV energy range, only $3\%$ belong to the elastic peak, and the remaining part is approximately uniformly distributed over the rest of the energy spectrum. This is the confirmation of the scenario proposed in \cite{ara05} of LEE presence in the gap between TM and EH. Moreover, being the TEY$>1$ in the energy interval $0.2-5$ keV (or more for higher incidence angles), Figure \ref{fig:yield} suggests that LEE would increase in number within the gap 
due to repeated electron emission occurring upon backscattering events at the interface of the TM and EH.

For studying quantitatively the impact of these electrons in the charging process of the LISA TM we must now turn to tests with a representative geometry of the TM and EH geometry considering the fluxes of primary particles characteristic of the space environment.
In this work we approach this study by using MC simulations based on GEANT4 toolkit \cite{ara05}.
The GEANT4 framework is a standard and widely used tool in particle physics that allows one to simulate particle propagation in matter. In particular, it is able to deal with the transport of elementary particles in complex geometries, including those of the LISA satellites, over a wide range of energies characterizing the galactic cosmic rays and high-energy solar particles. Therefore, this toolkit is fundamental for calculating the TM charging rate and charge noise on LISA TMs for a variety of environmental conditions in space, including both stationary and non-stationary particle fluxes. 

In the remaining part of the paper we first compare the LEE production in a GEANT4-based simulation with the results presented in Sect. \ref{s2}. This comparison represents an independent test of our ab initio calculations and sets the basis for the development of an end-to-end tool to simulate the TM charging effects on the LISA mission. The toolkit development is beyond the scope of the present work and will be discussed in a future paper. 
Finally, we provide a first quantitative analysis of the impact that the transport of LEE has in the TM charging simulation.

\subsection{GEANT4 Monte Carlo}

The GEANT4 engine allows for a modular management of the physical processes that can be added to the simulation in order to make it more precise at a cost of a higher CPU time. 
Such processes can be added in a so-called ``modular physics list''. Each process is implemented in terms of energy and material dependent cross sections which are modelled by different theoretical approaches that can be selected by the user.
This list is called ``modular'' as the software describing the physical processes of the same type (electromagnetic, hadronic, decay...) is collected in libraries that can be linked independently for proper simulations of different particles, energy ranges, and materials. 

Particularly important in our case is the electromagnetic library. A module within the library, initially developed for studying radiobiology and DNA damage by ionizing charged particles, is called GEANT4-DNA \cite{Sakata}. In its latest version, the module (formerly available only for water target to mimic human tissue) also includes the cross sections for solid gold. 
Having been designed to simulate the propagation of extremely low-energy electrons generated by ionizing radiation in biological materials and gold nano-particles, it provides reliable results down to electron energies of the order of $10$ eV. 
In general, a GEANT4 simulation allows for the propagation of each particle in discrete steps whose length is calculated from the highest cross section of the processes active in the region of the simulation geometry. To each of these steps, a continuous energy loss, proportional to the length of the step, is assigned depending on the material of the volume being crossed. This way of managing the ionization energy loss is very efficient in terms of computation time and is particularly reliable at high energies, when the step-lengths are relatively long. Conversely, this approach is not precise enough to describe accurately the energy spectrum of the simulated particles at extremely low energies, when the discrete nature of the energy loss process becomes important. Nevertheless, the number of secondaries produced and the deflections of the particle trajectories depend only on whether or not any inelastic interaction takes place and thus the GEANT4 simulation can be considered reliable in this respect.
This motivates the choice of TEY as the observable for comparing GEANT4 calculations with the SEED results described in Sect. \ref{s2}.

\subsection{Experiments and results}

An electron backscattering experiment was simulated by using the GEANT4 framework with a simplified experimental geometry  consisting of a crystalline, thick gold target and a mono-energetic electron beam of various energies ranging from $0.1$ to $50$ keV. The simulation includes a virtual detector counting every electron backscattered from the target, not distinguishing the  primary electrons from secondary electrons produced near the target surface escaping the material. This experiment is the exact reproduction of the one run with SEED in Sect. \ref{s2}. 

\begin{figure}[h!]
    \centering
    \includegraphics[width = 0.9\textwidth]{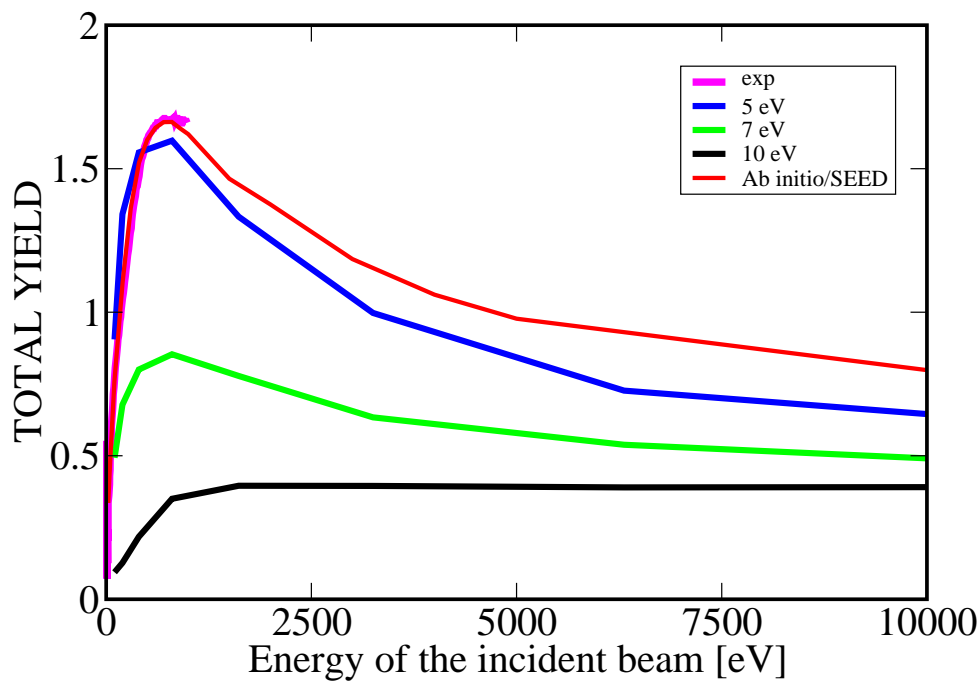}
    \caption{Total electron yields obtained by the GEANT4 simulation as a function of electron beam energy in comparison to experimental data (magenta line) \cite{gonzalez2017secondary} and SEED calculations (red line). The different curves show the results obtained by changing the minimum energy threshold for the secondary electron propagation, in particular: 5 eV (blue line), 7 eV (green line), and 10 eV (black line).}
    \label{fig:yield_th}
\end{figure}
In Figure \ref{fig:yield_th} the results of the simulated TEY as a function of the energy of the primary beam are presented. GEANT4-DNA manages the generation of low-energy electrons down to $10$ eV, but allows for the propagation of particles in the geometry at lower energies. The curves in Figure \ref{fig:yield_th} represent the yields obtained with different minimum propagation energies.


All the curves converge to a value of $\approx 0.5$ above approximately $10$ keV, in accordance with the expected backscattering probability of an energetic electron on a gold target. Below $10$ keV energies, the contribution of secondary electrons escaping the target becomes important, and the TEY increases even above $1$, but only if the minimum propagation  energy allows electrons to be propagated backward to the virtual detector.
The value of the minimum energy threshold that gives a result compatible with SEED calculation and experimental results is $\simeq 5$ eV \cite{gonzalez2017secondary} (see Figure \ref{fig:yield_th}). This result follows from the fact that $5$ eV is compatible with the gold electron work function used in Sect. \ref{s2}.

\begin{figure}[h!]
\centering
\includegraphics[width=0.5\textwidth]{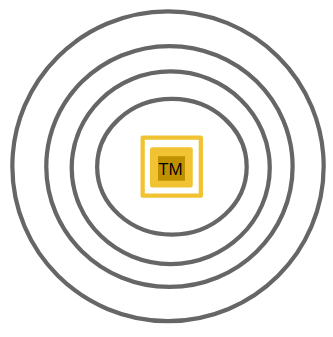}
\caption{Simplified geometry of the LISA TM system}
\label{fig:LISAGeom}
\end{figure}

\begin{figure}[h!]
\centering
\includegraphics[width=0.9\textwidth]{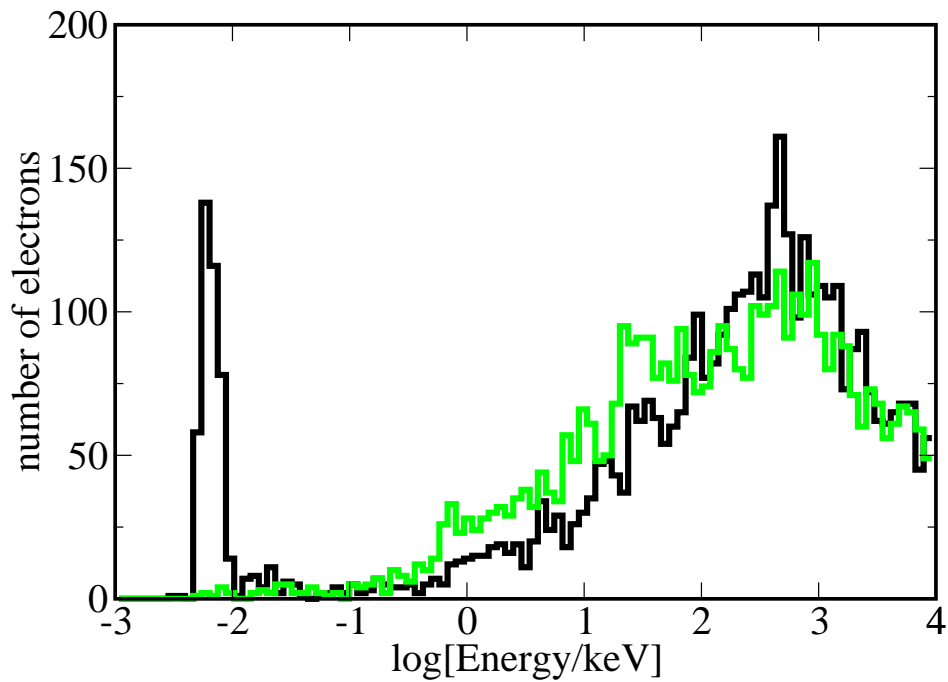}
\caption{Energy distribution of the electrons in the TM gap produced by a proton beam using LEE transport (GEANT4-DNA, black line) or adopting option 4 for the standard electromagnetic libraries (OPT4, green line)}
\label{fig:LEEspec}
\end{figure}

In order to highlight the importance of considering LEE in the simulation of the LISA TM charging, it is useful to provide a quantitative estimate. To this purpose we simulated a proton beam of $1$ GeV energy, representing the median energy of the galactic spectrum at $1$ AU, traversing $16\,\mathrm{g\,cm}^{-2}$ of material \cite{a&aub}. The geometry consists of four aluminum spherical shells enclosing a cubical box, $150$ nm thick and $5.2$ cm wide, representing the inner part of the EH, and the LISA cubic TM separated by a $3$ mm vacuum gap. This arrangement, sketched in Figure \ref{fig:LISAGeom}, is roughly representative of the material traversed by particles incident on the LISA spacecraft. The adoption of a simplified geometry is a common practice in the design process of space missions to carry out a first guess of the effects of the space environment on sensitive parts of the satellite \cite{telloni16}.
In the MC simulation, we selectively activate the LEE transport only on a gold $150$ nm thick layer on the TM and EH surfaces. This approach, described in \cite{bridging} allows us to save computation time without losing the significance of the simulation as the free mean path of $100$ eV electrons is of the order of $1$ nm in gold. 
Exactly the same experiment was simulated using the standard electromagnetic physics advised as default in GEANT4, available in different options. The option 4 (OPT4) of this set-up is able to model the electromagnetic processes for electrons down to $100$ eV \cite{Opt4}. 
Figure \ref{fig:LEEspec} shows the energy spectrum of the electrons intercepted by a virtual detector in the gap between the TM and the EH with LEE transport in comparison with OPT4.\\

The simulation with the LEE transport activated (black line in Figure \ref{fig:LEEspec}) produces with respect to the OPT4 one (green line in Figure \ref{fig:LEEspec}) an excess of electrons at the lowest (below $10\, \mathrm{eV}$) and the highest energies (close to $1\, \mathrm{MeV}$) and a reduction in the intermediate interval. This appears as a sort of redistribution of the electrons in the energy bins. However, by integrating the spectra, one would observe that the total number of the electrons produced with LEE is bigger by about $20-30\%$. Considering just the fraction up to $20$ eV, with LEE we count about one third of the total electrons and the increment is almost one order of magnitude. 
We stress that early MC simulations based on GEANT4 \cite{ara05,grim15,wass2005} systematically underestimated the TM charge noise in LPF because they were unable to account for these additional electrons.
We thus conclude that to properly estimate the TM charging during the LISA space mission is mandatory to use LEE in future MC simulations.
Moreover, we recall that local electric voltages of the order of $1$ V are applied by the sensing and actuation circuitry to control the TM motion in LISA as were in LPF. The resulting electric fields are big enough to influence the motion of the low-energy ($\approx 10$ eV) electrons\footnote{and also of all those electrons with small enough velocity parallel to the field} biasing the charging of the TM. This bias contribution can be estimated by using FEM-based software able to calculate the local electric fields in the gap between TM and EH from the timelines of the electrode commands and to trace the electrons paths.




\section{Conclusions}

In this work we shed light on the mechanisms that underlie the charging of the TMs hosted in drag-free spacecraft with particular reference to the LISA mission and its Pathfinder. We mainly focus on the role of the keV and sub-keV electrons aiming to accurately include their motion in the numerical modelling of the charge transport within the device at the interface between the TM and the EH.


We studied the electron transport within the gold TM by the in-house MC tool SEED to investigate in particular the low-energy regime. SEED uses as input the elastic and inelastic cross sections of Au. The required inelastic mean free path was calculated numerically using the dielectric approach, which is based on the ELF. The ELF of Au was assessed from first principles calculations up to $70 $ eV and extended to include higher single-particle excitation energies by matching the ab-initio results with the experimental NIST X-ray atomic data. The elastic mean free path to model the trajectory deviation of the electrons scattered out by the atomic nuclei of the target was calculated via the Mott theory. 

Using our MC approach, we were able to follow the trajectories, also in their way out of the solid, of the backscattered primary beam as well as of the secondary electrons emerging from the surface at very low energy. This information has been used to calculate i) the energy spectrum of the electrons down to very low energy and ii) the TEY in the $0-10000$ eV energy range for different angles of incidence.

The spectral lineshape reveals a crucial role in the TM charging of the electrons below $50$ eV, that was discussed, but not included in previous MC simulations concerning the LISA satellite. Moreover, we used these results to customize a framework based on the GEANT4 11.0 toolkit in which simulations can include the contribution of electrons down to $5$ eV. In this set up, we were able to reproduce the TEY calculated using SEED and to give a first estimate of the contribution of the LEE in a relevant test-case for the LISA TMs.
To this purpose, we considered a $1$ GeV proton flux, representing the average energy of the galactic cosmic ray spectrum at $1$ AU at the solar minimum and a simplified LISA spacecraft geometry. We have found that the low-energy electron transport raises by $20-30\%$ the total number of electrons (secondary and backscattered) contributing to the TM charging. A consistent fraction (one third) of the total number of electrons is below $20$ eV and would be able to bias the TM charging under the influence of the low voltages applied to the sensing and actuation electrodes in the LISA science operations.

\section*{Appendix (Calculation of the macroscopic dielectric function)}

To calculate the macroscopic dielectric function, we start from the polarization (or density-density response) function $\chi({\bf{r},\bf{r'}},W)$
\begin{equation}
\rho^{ind}({\bf{r}},W)=\int d{\bf{r'}}\chi({\bf{r},\bf{r'}},W)V^{\mathrm{ext}}({\bf{r'}},W) \, \mbox{,}
\end{equation}
where $\rho^{ind}({\bf{r}},W)$ is the electron density induced by the external Coulomb potential $V^{\mathrm{ext}}$, and $W$ is the electron excitation energy (or the electron energy loss).
$\chi({\bf{r},\bf{r'}},W)$ can be directly related to the electronic band structure of the material. For periodic solids, one can use one-electron Bloch states expanded in terms of plane-waves as follows:
\begin{equation}
\phi_{{\bf q},n}({\bf r})=\frac{1}{\sqrt{V}}\sum_{\bf G} u_{{\bf q},n}({\bf G})\exp^{i({\bf q}+{\bf G})\cdot {\bf r}},
\end{equation}
where $V$ is the volume of the simulation cell, ${\bf q}$ is the electron momentum (or momentum transfer) within the 1BZ, ${\bf G}$ is a reciprocal lattice vector, $n$ identifies the band, and $u_{{\bf q},n}$ are functions with the same periodicity as the crystal lattice.
For periodic crystals, like bulk Au, the polarizability $\chi({\bf{r},\bf{r'}},W)$ can be conveniently expressed by Fourier transform to the reciprocal space as a matrix  $\chi_{\mathbf{G},\mathbf{G'}}({\bf q},W)$, and obtained by solving the following Dyson equation \cite{RevModPhys.74.601, Weissker2010}:
\begin{eqnarray}
\chi_{\mathbf{G},\mathbf{G'}}({\bf q},W) = \chi^0_{\mathbf{G},\mathbf{G'}}({\bf q},W)+ \sum_{\mathbf{G''},\mathbf{G'''}}\chi^0_{\mathbf{G},\mathbf{G''}}({\bf q},W) \nonumber \\ \times \Big[ v_{\mathbf{G''}}({\bf q})\delta_{\mathbf{G''},\mathbf{G'''}}+ K^{xc}_{\mathbf{G''},\mathbf{G'''}}({\bf q},W) \Big]\chi_{\mathbf{G'''},\mathbf{G'}}({\bf q},W),
\end{eqnarray}
where $\mathbf{G'},\mathbf{G''},\mathbf{G'''}$ are reciprocal lattice vectors, $v_{\mathbf{G'}}({\bf q})=4\pi/|{\bf q}+{\bf G'}|^2$ is the Fourier-transformed bare Coulomb potential, and 
$\chi^0_{\mathbf{G},\mathbf{G'}}({\bf q},W)$ is the non-interacting polarization function that is reckoned by knowing the Kohn-Sham eigensolutions; finally,
$K^{xc}_{\mathbf{G''},\mathbf{G'''}}({\bf q},W)=\left\{ \frac{d}{d\rho}v_{\rm  xc}[\rho]\right\}_{\rho=\rho(\textbf{r},t)}$ is the time-dependent density functional theory (TDDFT) energy- and momentum-dependent kernel, which accounts for the exchange-correlation interaction and has been modeled via the adiabatic local density approximation (ALDA),
where $\rho$ is the DFT ground state density local in both time and space, and $v_{\rm  xc}[\rho]$ is the local density approximation (LDA) to the (generally unknown) exchange-correlation functional.
From the knowledge of ${\chi_{\mathbf{G},\mathbf{G'}}}({\bf q},W)$ the microscopic dielectric function reads:
\begin{equation}
\epsilon_{\mathbf{G},\mathbf{G'}}({\bf q},W)=1+v_{\bf G'}({\bf q})\chi_{\mathbf{G},\mathbf{G'}}({\bf q},W) \, \mbox{,}
\end{equation}
where $1=\delta_{\mathbf{G},\mathbf{G'}}$ is the identity operator. In SEED $\mathbf{q}$ represents the transferred momentum vector within the 1BZ.

Finally, the macroscopic $\bar{\epsilon}( \mathbf{q}, W)$ and microscopic $\epsilon_{\mathbf{G},\mathbf{G'}}(\mathbf{q}, W)$ dielectric functions are related by \cite{Wiser}:
\begin{equation}\label{eq:eM}
\bar{\epsilon}( \mathbf{q}, W) =  \left [ \epsilon^{-1}_{\mathbf{G}=0,\mathbf{G'}=0}(\mathbf{q}, W)  \right ]^{-1}.
\end{equation}
The inversion of the full dielectric matrix before taking the head element of the matrix allows one to include the crystalline local field effects \cite{doi:10.1119/1.12734}, which account for the local anisotropy of the material. These effects are deemed to be important for Au, particularly above 35 eV \cite{GURTUBAY2001123,PhysRevB.88.195124}.

\section*{Acknowledgments}
This action has received funding from the European Union under grant agreement n. 101046651 and from the European Space Agency under ESA Contract No. 4000133571/20/NL/CRS.
\section*{References}
\bibliographystyle{unsrt}
\bibliography{sample.bib}
\end{document}